%% file: ms.tex
\documentclass[12pt,preprint]{aastex}

\newcommand{\teff}{T$_{\rm eff}$} 
\newcommand{\hipparcos}{{\scshape{Hipparcos}} }

\begin{document}

\title{A search for cool subdwarfs: Stellar parameters 
for 134 candidates\footnote{Based in-part on observations obtained 
with the Hobby-Eberly Telescope, which is a joint project 
of the University of Texas at Austin, the Pennsylvania State University, 
Stanford University, Ludwig-Maximillians-Universit\"at M\"unchen, and
Georg-August-Universit\"at G\"ottingen.}}

\author{David Yong and David L. Lambert}

\affil{Department of Astronomy, University of Texas, Austin, TX 78712}
\email{tofu,dll@astro.as.utexas.edu}

\begin{abstract}

The results of a search for cool subdwarfs are presented.
Kinematic (U, V, and W) and stellar parameters 
(\teff, log g, [Fe/H], and $\xi_t$) are derived for 
134 candidate subdwarfs based on high resolution 
spectra.  The observed stars span 4200K $<$ \teff~$<$ 6400K and
$-2.70<$ [Fe/H] $<$ 0.25 including only 8 giants 
(log g $<$ 4.0).  Of the sample, 100 stars have MgH bands
present in their spectra.  The targets were selected by their
large reduced proper-motion, the offset from the solar metallicity 
main sequence, or culled from the literature.  We confirm the claims
made by \citet{ryan89} regarding the NLTT catalog being a rich
source of subdwarfs and verify the success of the reduced proper-motion
constraint in identifying metal-poor stars.

\end{abstract}


\section{Introduction}
\label{sec:intro}

The driving force behind our understanding of the chemical evolution
of the Galaxy is the interpretation of observed abundance
ratios.  Self-consistent analyses of large high quality 
data sets have revealed detailed abundance patterns (e.g., \citealt{bdp93},
\citealt{mcwilliam95}).  Recent attempts
to understand the observed abundance trends include endeavors by
\citet{timmes95}, \citet{goswami00}, and \citet{alc01}
who predict the evolution of the 
abundances of all elements from carbon through zinc. 
Cool stars provide a unique opportunity to test directly these
models of Galactic chemical evolution through the
abundances of low ionization potential trace
elements (e.g., Rb, Cs) and isotopic ratios measured from
molecular bands (e.g., Mg from MgH, Ti from TiO).  
Presently these tests cannot be carried out due to the dearth of known
cool metal-poor stars.  In order to rectify this situation we have
commenced a search for cool subdwarfs, unevolved metal-poor stars
that fall below the solar metallicity main sequence in color-magnitude
diagrams.

Searches undertaken by \citet{carney87} and 
\citet{ryan89} selected candidates drawn from
proper-motion catalogs.  These searches were successful in
identifying metal-poor dwarfs for subsequent detailed abundance analysis.
However, both searches neglected cool subdwarfs.
An alternative to searches for metal-poor stars based on proper-motion
catalogues are the objective prism surveys which 
identify candidate metal-poor stars by the
weakening of Ca {\scshape ii} H and K. 
Such studies have been conducted by \citet{bond70},
\citet{bond80}, \citet{beers85}, \citet{beers92}, and others.
Selection criteria which rely exclusively upon Ca features as metallicity
indicators strongly bias the sample towards warmer stars.
Since metal lines 
weaken with increasing \teff, a temperature
estimate is required before deciding if lines are abnormally weak.
In the absence of temperature indicators, a cool
metal-poor star will have Ca {\scshape ii} features comparable to a warmer
solar metallicity star.
More current work by \citet{christlieb00} on the
stellar content of the Hamburg/ESO Survey has dramatically
increased the yields of metal-poor stars.  Candidate metal-poor stars are
identified from their location in {\em $(B-V)$ $-$ Ca line strength} space.  
By comparing the Ca line strength between stars of similar \teff's,
this increases the likelihood that a metal deficiency is the
cause of the relative weakening of the Ca line in a candidate.
With the remarkable success rate of 
60\% for stars below [Fe/H] $=-2.0$, twice as good
as the HK survey by Beers and collaborators, we look forward to 
the results of the Hamburg/ESO Survey when applied to cool subdwarfs.

In this paper we present the results of our search for cool subdwarfs, 
stellar (\teff, log g, [Fe/H], and $\xi_t$) and kinematic
parameters (U, V, and W) for 134 stars.
Our sample includes 80 stars with no prior
metallicity estimates.  In \S\ref{sec:criteria} we
outline and justify the selection criteria.  The observations will be
described in \S\ref{sec:data} and the analysis 
in \S\ref{sec:analysis}.  A discussion will be presented in
\S\ref{sec:discussion} and concluding remarks given in
\S\ref{sec:remarks}.

\section{Selection criteria}
\label{sec:criteria}

Our goal was to find previously unidentified subdwarfs with 
${\rm 4000K}<$ \teff~$<{\rm 4700K}$ and ${\rm [Fe/H]}<-1.5$.
We modelled our search upon the highly successful effort by \citet{ryan89}
using the NLTT catalog of stars with annual proper-motions in excess
of $0\farcs18$/year \citep{luyten79,luyten80}.
Subdwarfs are on plunging orbits, often highly elliptical, resulting in large transverse
velocities relative to the local standard of rest.  Subdwarfs are therefore 
over-represented in proper-motion catalogues.  
Following \citet{ryan89}, our primary criterion
for ensuring that metal-poor candidates were selected was the
reduced proper-motion, $H=m_V+5\log\mu+5$ where 
$m_V$ is the apparent magnitude and
$\mu$ is the proper-motion in arcsec/year.  The reduced proper-motion
can also be expressed as $H=M_V+5\log\nu_T-3.37$ where $M_V$ is
the absolute magnitude and $\nu_T$ is the transverse velocity in
km/s.  At a given spectral type, a reduced proper-motion constraint
rejects stars whose transverse velocities fall below a chosen
value.  Subdwarfs are less luminous than disk dwarfs at 
a given color and have larger transverse velocities.  Both effects
conspire to increase the reduced proper-motion of the subdwarf population
relative to the disk dwarf population.
We adopted Ryan's reduced proper-motion
constraint requiring stars have
$H_R \ge 10.7 + 2(m_{\rm pg}-m_R)$ where $H_R=m_R+5\log{\mu}+5$ using
Luyten magnitudes $m_R$ and $m_{\rm pg}$.
We imposed a magnitude limit $m_R \le 13.0$ and a declination limit
$-40^o \le \delta \le 90^o$ as our observations were made 
at McDonald Observatory where we
required reasonable signal-to-noise (60 per pixel at 6500\AA) in
20-30 minute exposures.  To ensure that we observed cool stars,
Luyten color classes g-k, k, and k-m were selected.  
\citet{carney87} and \citet{ryan89} previously observed the
g-k color class.  We included these stars so that we could
compare our derived stellar parameters 
with the Carney and Ryan values.  Ryan's
photometry suggested that the color classes were not
accurate.  Stars assigned color class g-k by Luyten
actually span late F to late K.  Further, the effect of 
the ultraviolet excess, the hallmark of a subdwarf, is to 
camouflage a cool subdwarf as a warmer disk dwarf in the Luyten
color system.  Altogether, our selection criteria resulted
in 4445 NLTT candidate subdwarfs.  

Our list was augmented by subdwarfs previously identified
in the literature.  Our final sources of candidate subdwarfs
were color-magnitude diagrams constructed 
using the \hipparcos \citep{hipparcos}
and Yale \citep{yale95} parallaxes.  Targets located beneath the solar metallicity
main sequence in the range $0.8 < B-V < 1.4$ were selected.
Subdwarfs have a higher \teff~than disk dwarfs at a given mass
which shifts subdwarfs blueward of the disk main sequence.
\citet{reid01} demonstrated that in the range
$B-V<0.8$, the majority of stars below the disk main sequence in
a color-magnitude diagram based on \hipparcos data
are not subdwarfs, [Fe/H] $>-1.0$.  Errors in the colors 
cause these disk dwarfs to appear as subdwarfs.  Our observations and analysis will 
test if the \citet{reid01} findings are valid when we extend 
to cooler dwarfs, $0.8 < B-V < 1.4$.  In total, we observed around
230 candidate subdwarfs though in this paper we are reporting
on 134 candidates.

\section{Observations and data reduction}
\label{sec:data}

\begin{figure}
\epsscale{0.9}
\plotone{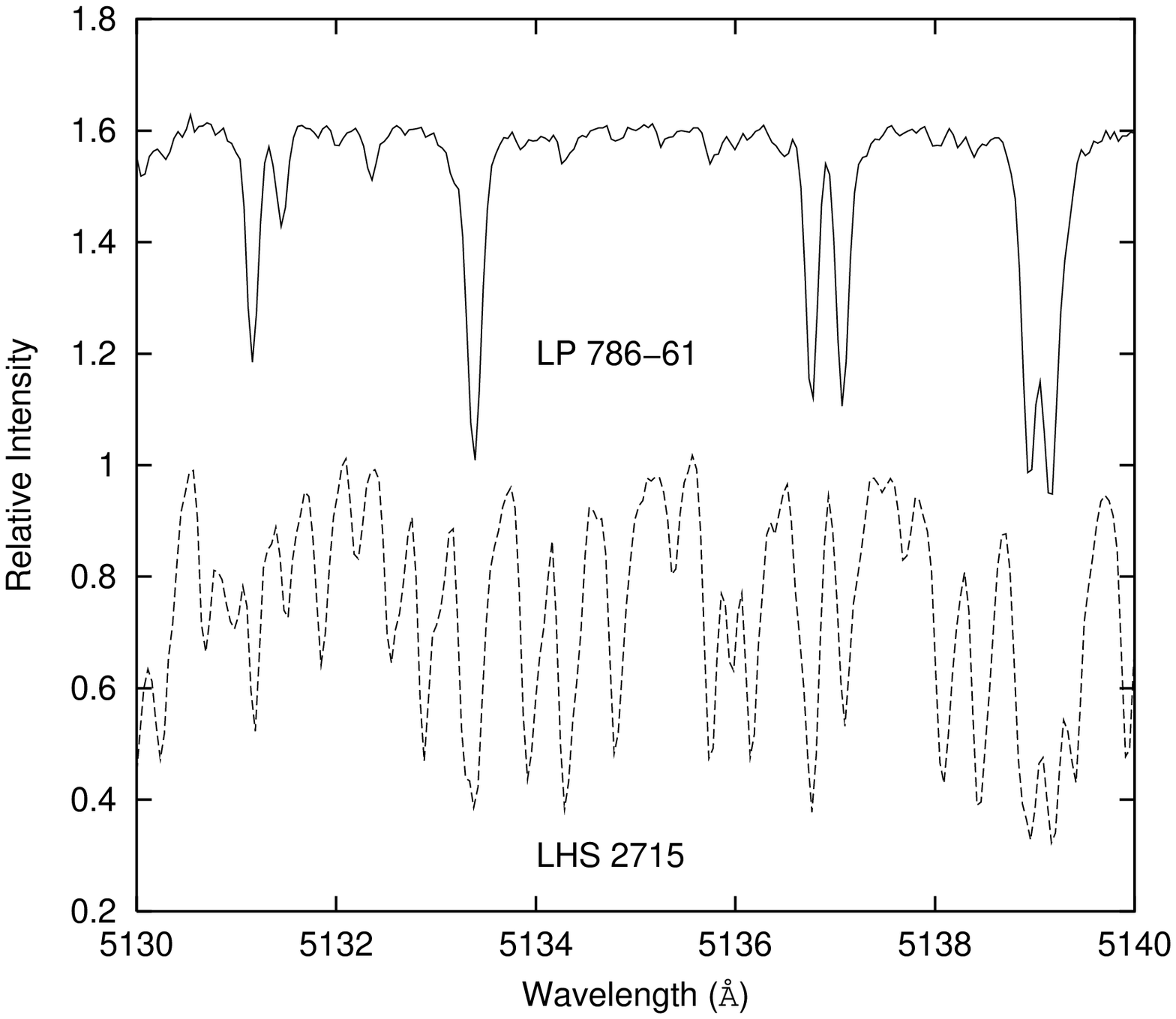}
\caption{R=60,000 data showing the absence (upper) 
and presence (lower) of the MgH A-X lines near 5140\AA.  
Many of the additional lines in the lower spectrum are due to
MgH.  \label{mgh}}
\end{figure}

\begin{figure}
\epsscale{0.9}
\plotone{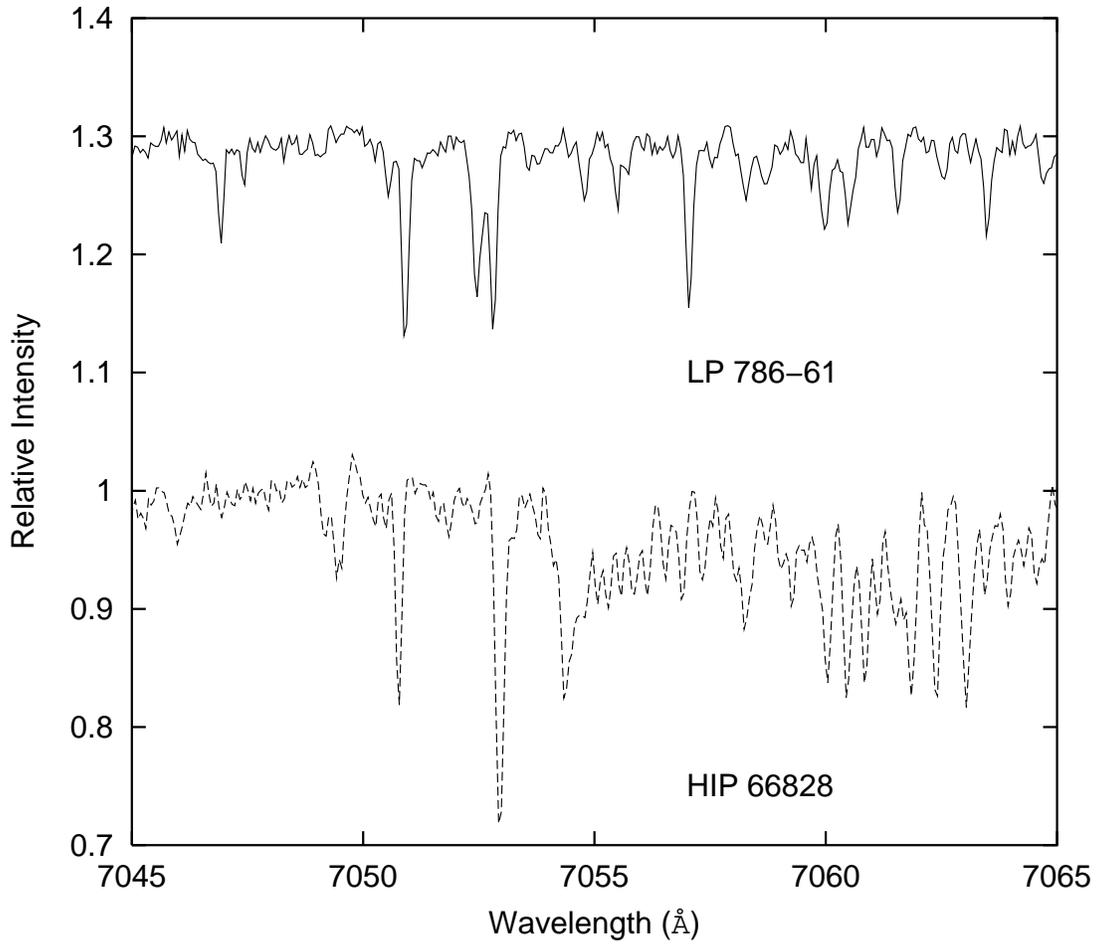}
\caption{R=60,000 data showing the absence (upper) 
and presence (lower) of the TiO $\gamma$-system 
A$^3\Phi-$X$^3\Delta$ bandhead near 7054\AA. \label{tio}}
\end{figure}

\begin{figure}
\epsscale{0.8}
\plotone{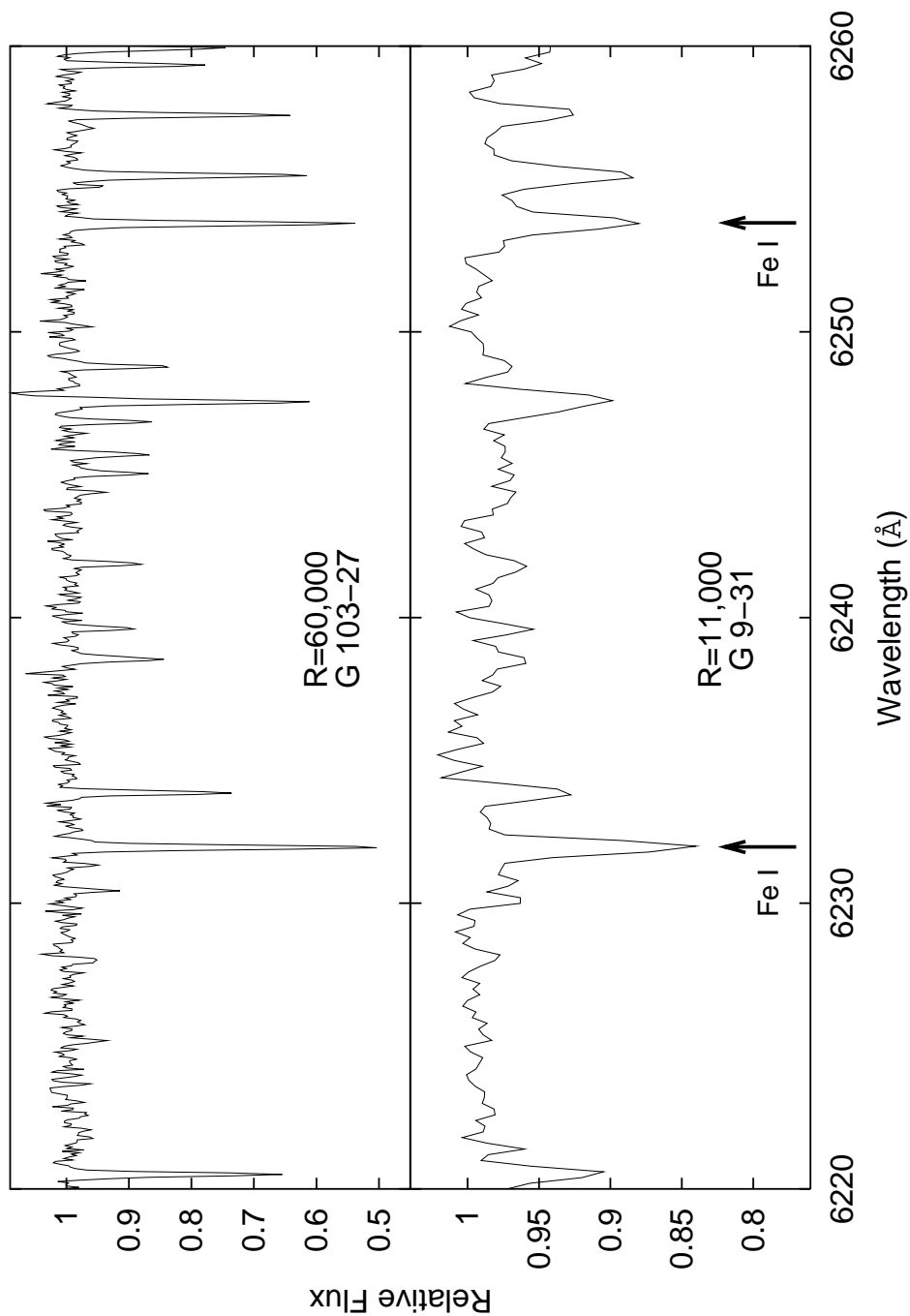}
\caption{Spectra of G 103-27 taken at R=60,000 and G 9-31 take at 
R=11,000.  Both stars have similar stellar parameters.
Representative Fe {\scshape i} lines used to derive stellar parameters are 
highlighted.  Importantly, the R=11,000 spectra has Fe {\scshape i} lines
for which equivalent widths can be measured.  \label{resolve}}
\end{figure}

Table \ref{param} contains the list of candidates observed
at McDonald Observatory on the 2.7m Harlan J. Smith telescope
and on the 9.2m Hobby-Eberly telescope (HET) 
between November 1999 and April 2002.  The 2.7m data were obtained
using the cross-dispersed echelle spectrometer \citep{tull95}
at the coud\'{e}~f/32.5 focus with a resolving power of either
30,000 or 60,000.  The detector was a 
Tektronix CCD with 24 $\mu{\rm m}^2$ pixels
in a $2048 \times 2048$ format.  For this setting, the 
spectral coverage was from 3800\AA~to 
8900\AA~with gaps between the orders beyond 5800\AA.  The HET data were taken with the
Upgraded Fiber Optic Echelle spectrograph \citep{harlow96} at a resolving power of
11,000 on a $1024 \times 1024$ CCD.  The spectral 
coverage was from 4500\AA~to 9000\AA~with gaps between the
orders beyond 7300\AA.  For observations on both 
telescopes, wavelength coverage incorporated the
MgH A-X bands near 5140\AA~as well as the TiO (0,0) bandhead of the
$\gamma$-system A$^3\Phi-$X$^3\Delta$ near 7054\AA.  Visual inspection
of the spectra readily indicated the presence or absence of the MgH or TiO molecular
features (see Figures \ref{mgh} and \ref{tio}).  Of our sample of 134 stars, 
34 displayed neither MgH nor TiO bands and 100 showed MgH or MgH and TiO bands.
Numerous Fe {\scshape i}, Fe {\scshape ii}, Ti {\scshape i}, 
and Ti {\scshape ii} lines were available in the observed
spectra for spectroscopic determination of stellar parameters.
For each star, exposure times were
generally 20-30 minutes and only occasionally were multiple
exposures taken and co-added.
Although varying from star to star, the typical signal-to-noise
of the extracted one dimensional spectra was 60 per pixel at 6500\AA.
One dimensional wavelength calibrated normalized spectra
were extracted in the standard way using the
IRAF\footnote{IRAF is distributed by the National Optical Astronomy Observatories,
which are operated by the Association of Universities for Research
in Astronomy, Inc., under cooperative agreement with the National
Science Foundation.} package of programs.  Equivalent widths were measured
using IRAF where in general Gaussian profiles were fit to the observed profile.
Figure \ref{resolve} shows the spectra of 2 different stars, one taken with R=60,000
and the other at R=11,000 where two Fe {\scshape i} lines used in the analysis are indicated.
Although less than ideal, an equivalent width analysis of R=11,000 spectra is feasible.

\section{Analysis}
\label{sec:analysis}

\subsection{Deriving stellar parameters}

\begin{figure}
\epsscale{0.8}
\plotone{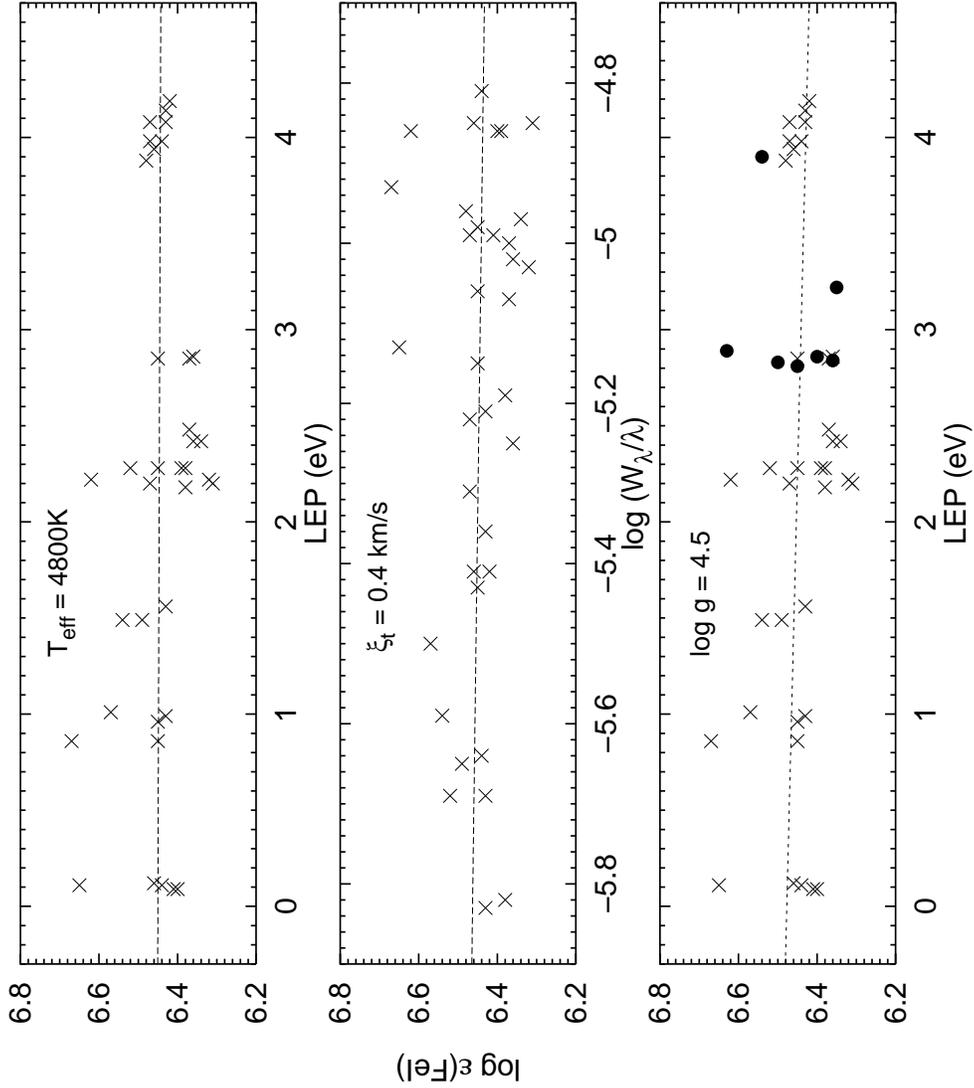}
\caption{Determination of stellar parameters \teff, log g, and $\xi_t$ using
excitation and ionization equilibrium for GJ 1064B.  In the top panel, the lower excitation
potential (LEP)-abundance relation is used to set \teff.  In the middle panel,
the reduced equivalent width (W$_\lambda/\lambda$)-abundance relation is used
to determine $\xi_t$.  In the bottom panel, the abundances of Fe {\scshape i} (crosses)
and Fe {\scshape ii} (filled circles) are used to fix log g.  In all panels the line
represents the least squares fit to the data. \label{fparam}}
\end{figure}

The LTE stellar line analysis 
program {\scshape Moog} \citep{sneden73} was used in
combination with the adopted model atmosphere.
For log g $>$ 3.5, we used 
the {\scshape Nextgen} model atmosphere grid for
low mass stars computed by \citet{nextgen99}.  For giants with
log g $\le$ 3.5 we used LTE model atmospheres computed by \citet{kurucz93}.
In both cases we interpolated within the grid when necessary.
We derived the model parameters in the following way.  We measured the equivalent widths
of $\sim$ 35 Fe {\scshape i} and $\sim$ 5 Fe {\scshape ii} lines.  Values for \teff~were 
set from excitation equilibrium.  
\teff~was adusted so that Fe abundances derived from individual lines were
independent of excitation potential, as they must be.
The Fe abundances derived from individual Fe {\scshape i} lines must
be independent of the strength of the line.  Thus the microturbulence, $\xi_t$, was
estimated.  
Finally, the gravity was determined from ionization equilibrium.  That is, the
Fe abundance derived from Fe {\scshape i} lines must agree with the abundance
derived from Fe {\scshape ii} lines.
This process was iterated until a consistent set of
parameters were obtained (see Figure \ref{fparam}).  Ti {\scshape i} 
and Ti {\scshape ii} lines were also used to check the stellar
parameters.  The final value of [Fe/H] was simply the weighted
mean of the Fe {\scshape I} and Fe {\scshape II} abundances derived from the accepted
model (see Table \ref{param} for model parameters).

The $gf$-values of the Fe lines used in the analysis were selected from \citet{lambert96}
and a list compiled by R.~E. Luck (1993, private communication).  
The $gf$-values of the Ti lines were selected from the R.~E. Luck 
compilation.  Where possible, only
weak lines, EWs $<$ 90 m\AA, were used in the analysis.  We checked
our analysis techniques by observing the solar spectrum at R=60,000.  We measured
30 Fe {\scshape i} lines and 7 Fe {\scshape ii} lines and 
compared our equivalent widths with the \citet{grevesse99} values.
Our equivalent widths were larger by a mean value of 3.7 m\AA~with a standard deviation
of 2.6 m\AA.  Using our equivalent widths and a {\scshape Nextgen} model atmosphere, we derived
a solar abundance of log $\epsilon$ (Fe) $=7.54$.  Considering the \citet{grevesse99}
value of log $\epsilon$ (Fe) $=7.50 \pm 0.05$ derived from their empirical model solar
atmosphere, we adopted log $\epsilon$ (Fe) $=7.52$
as the solar Fe abundance for this study.

Our derived model parameters, \teff, log g,
$\xi_t$, and [Fe/H] have associated uncertainties.  
We varied \teff~until the  trends between lower
excitation potential and abundance were unacceptable.
Similarly, we took values for $\xi_t$ that noticeably changed
the trends between equivalent width and abundance.
For log g, we measured the
standard deviation of the Fe abundance as derived from Fe {\scshape i} lines 
(typically 0.1 to 0.15 dex) then
allowed Fe {\scshape i} and Fe {\scshape ii} to 
agree within this standard deviation.  This produced
uncertainties of $\delta$\teff=150K, $\delta$log g=0.3 dex, 
$\delta\xi_t$=0.3 km/s, and $\delta$[Fe/H]=0.2 dex in the model
parameters.

\subsection{Comparison with literature}

\begin{figure}
\epsscale{0.9}
\plotone{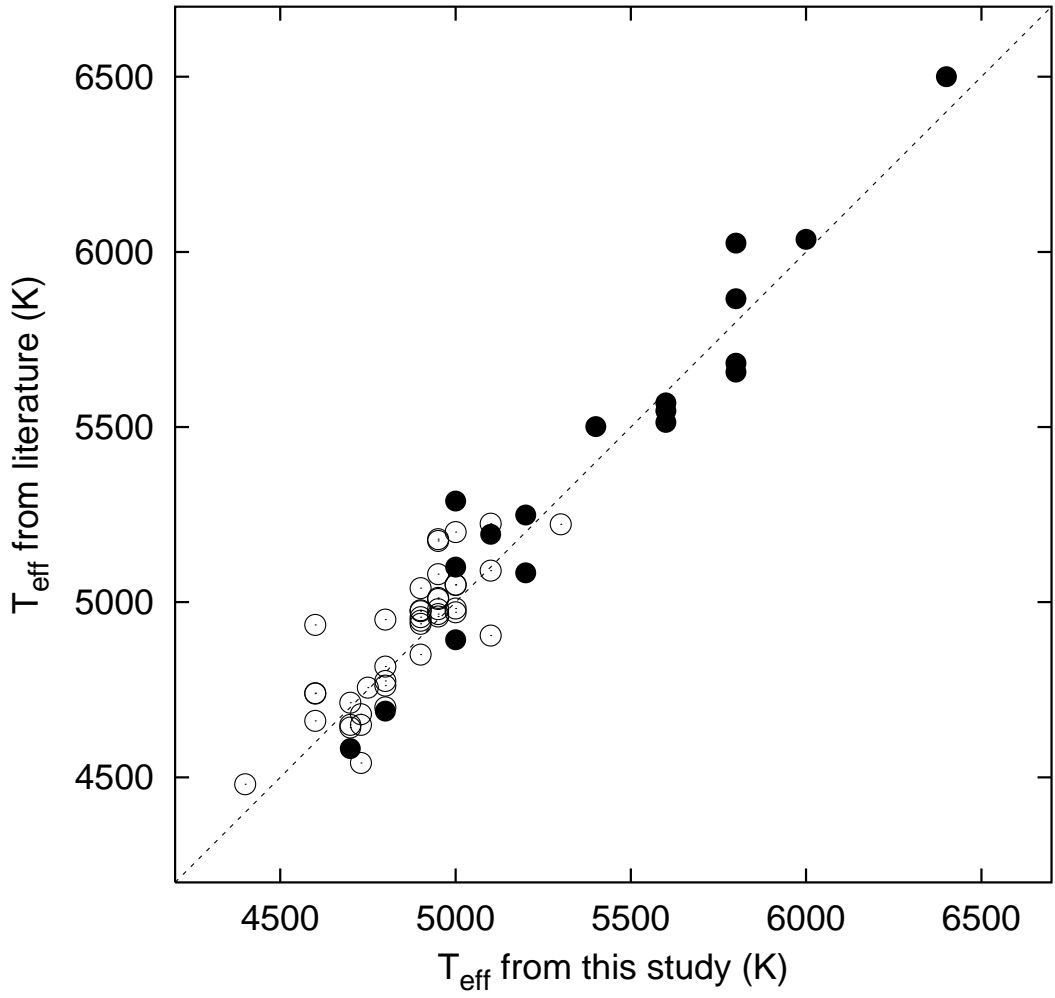}
\caption{\teff~comparisons between this work and various literature studies.  The
filled circles represent stars which have neither MgH
nor TiO bands in their spectra.  The dotted line represents the line of
equality. \label{teff}}
\end{figure}

\begin{figure}
\epsscale{0.9}
\plotone{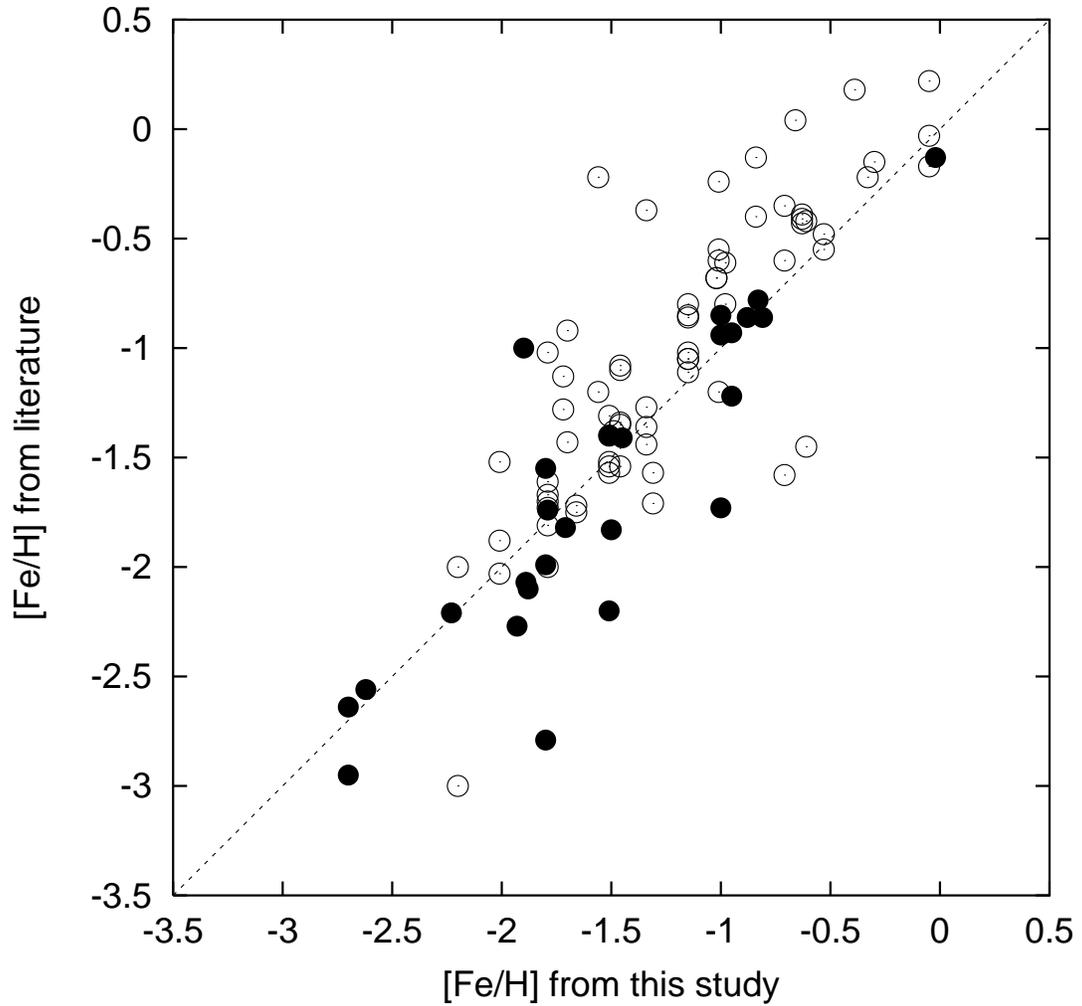}
\caption{[Fe/H] comparisons between this work and various literature studies.  The 
filled circles represent stars which have neither 
MgH nor TiO bands in their spectra.  The dotted line represents the line of
equality.  We note that the scatter for the open circles is slightly larger
than for the filled circles.
\label{feh}}
\end{figure}

For the 134 candidate subdwarfs analyzed here, a search on SIMBAD
indicated that there were 37 stars
with previously determined values for \teff~and 54 stars with
previously determined values for [Fe/H].  We have compared our values
with those found in the literature (see Table \ref{comp} and Figures \ref{teff} and \ref{feh}).
We find a mean offset $\langle$\teff (this study)$-$\teff (literature)$\rangle$ 
$= -25$ K with a standard deviation of 114K.  For [Fe/H], the mean offset
is $\langle$[Fe/H] (this study)$-$[Fe/H] (literature)$\rangle = -0.1$ dex
with a standard deviation of 0.38 dex.  The agreement is reasonable 
between the stellar parameters derived in this study and the
values found in a variety of sources in the literature.

Table \ref{comp} shows values for [Fe/H] given by \citet{ryan91}
for a number of stars.  However, these metallicities are not paired with
values for \teff.  
The \citet{ryan91} sample are re-observations of a subset of metal-weak stars
identified in \citet{ryan89}.  An index measuring the
strength of the Ca {\scshape ii} K line was calibrated empirically as a function of
$B-V$ to give [Fe/H]. 
\citet{alonso96,alonso99} determined \teff's using the
infrared flux method which requires a metallicity and gravity estimate.  The 
derived \teff's are insensitive to the input metallicity and gravity.
We also computed \teff's derived by use of the {\em \teff:[Fe/H]:color} relations
given in \citet{alonso96b,alonso99b} assuming values for [Fe/H] from this
study, see Table \ref{irfm}.

MgH and TiO bands will not be visible in the spectra of stars
with sufficiently high temperatures and/or low abundances.  
Indeed, this is the case for 34 of the stars as noted in the last column of 
Table \ref{param}.

\section{Discussion}
\label{sec:discussion}

\subsection{Kinematics}

\begin{figure}
\epsscale{0.9}
\plotone{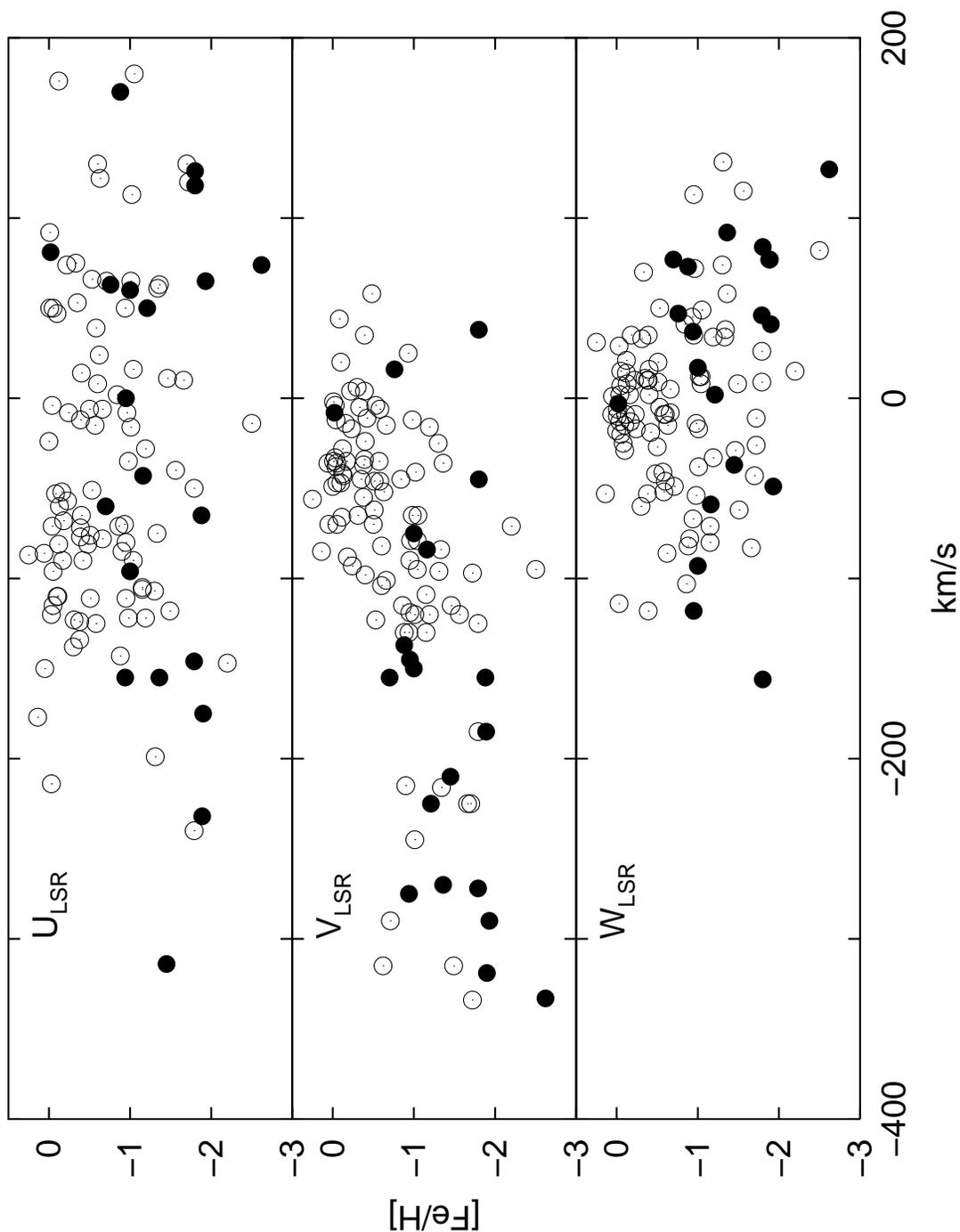}
\caption{Galactic space-velocity U,V, and W versus [Fe/H] where U,V, and W are relative
to the local standard of rest (LSR).  Filled circles represent stars which have neither 
MgH nor TiO bands in their spectra.  Only stars which have
$\frac{\sigma_U+\sigma_V+\sigma_W}{|U|+|V|+|W|}<0.7$ are 
shown, 117 out of 134.  A considerable fraction
of the sample noticeably lag the LSR (V $<-50$ km/s) \label{uvw}}
\end{figure}

\begin{figure}
\epsscale{0.8}
\plotone{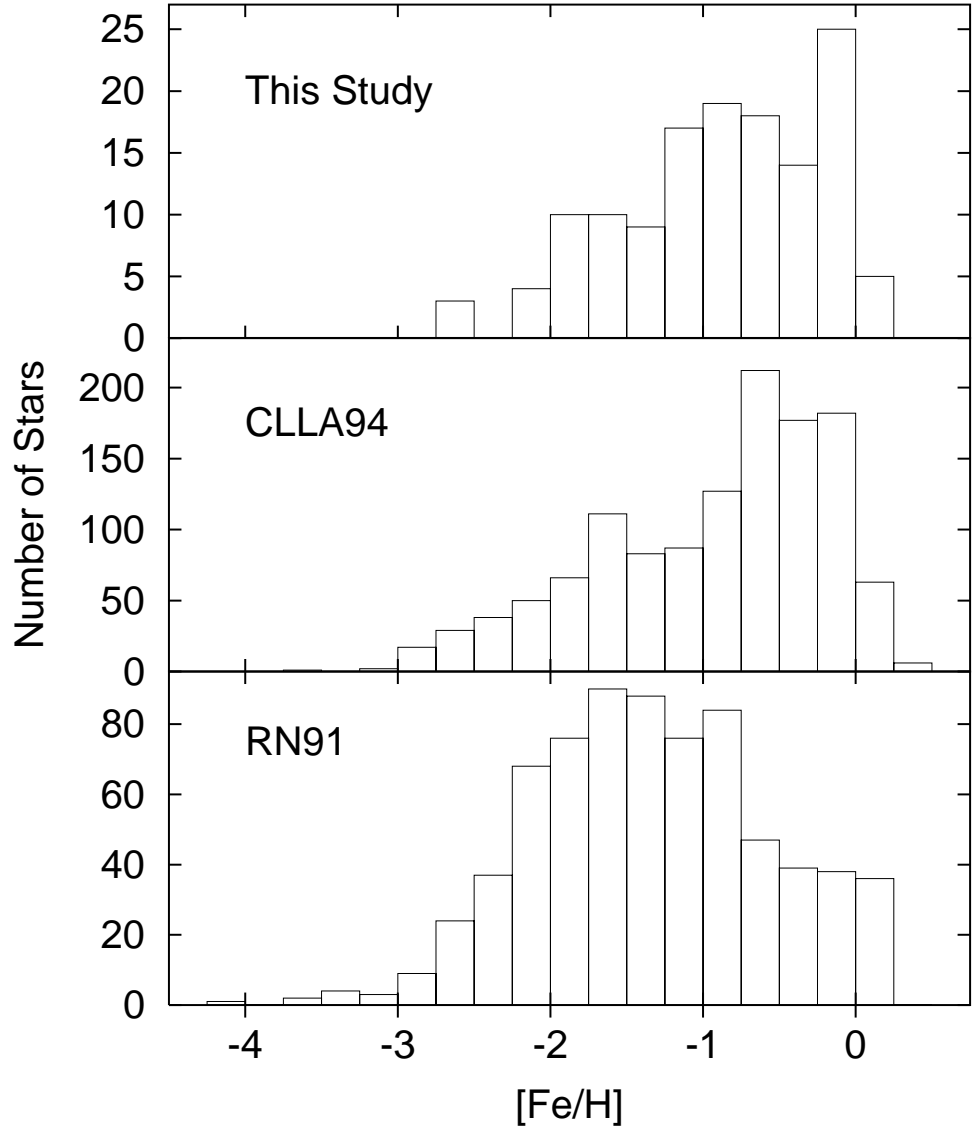}
\caption{Number of stars versus metallicity for this study (upper), 
the \citet{carney94} study (middle), and the \citet{ryan91} study (lower). 
Note the similar distributions between this study
and the Carney work.  The Ryan distribution peaks at a lower
metallicity since their sample consists of stars identified as having an ultraviolet
excess corresponding to [Fe/H] $< -1.2$. \label{histogram_fe}}
\end{figure}

We selected stars with large reduced proper-motions and 
therefore expect the observed sample to be kinematically distinct from
the thin disk.  
Following the prescription given by \citet{johnson87}, the
Galactic space-velocity components U, V, and W were calculated along
with the associated uncertainties (see Table \ref{param}).  To correct
for the solar motion with respect to the local 
standard of rest (LSR) we assumed the \citet{lsr} values 
($-$10,+5,+7) km s$^{-1}$ in (U,V,W).  
In the absence of \hipparcos parallaxes, 
spectroscopic parallaxes were determined by
using the derived model parameters and the \citet{girardi2000} isochrones.
Figure \ref{uvw} shows [Fe/H] versus U$_{LSR}$, 
V$_{LSR}$, and W$_{LSR}$.  We identify stars which
lag the LSR, V $<-$50 km/s, as being members of the thick disk or 
halo.  As expected from a reduced proper-motion constraint, we 
have selected stars which belong to populations kinematically distinct
from the thin disk.  
Many of the observed stars have abundances indicative of the thick disk and halo.
\citet{ryan89} claimed that the NLTT catalog was a rich source
of subdwarfs as identified by the ultraviolet excess.  Figure \ref{histogram_fe}
shows the numbers of stars for various metallicity bins.  Given
that 27 stars have [Fe/H] $\le -1.5$, our data endorse
Ryan's claims.  From the kinematics and abundances of the observed 
stars, we conclude that the reduced proper-motion constraint
successfully selected metal-weak thick disk and halo stars.

\subsection{Temperature}

\begin{figure}
\epsscale{1.0}
\plotone{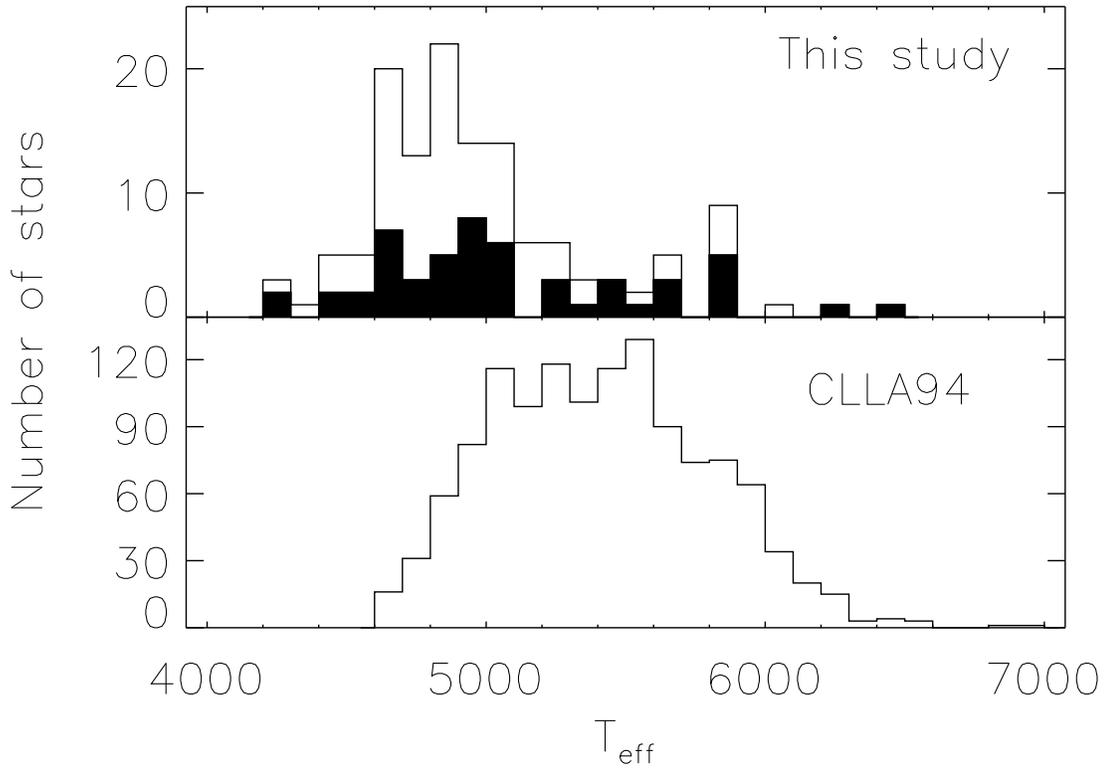}
\caption{Number of stars versus \teff~for this study (upper) and
the \citet{carney94} study (lower).  In the upper panel, the
filled in histogram represents the distribution of stars with
[Fe/H] $\le -1.0$.
The distribution for our
sample peaks at lower values of \teff~than the Carney
sample.  This was expected since we deliberately 
selected cooler color classes.  \label{histogram_teff}}
\end{figure}

Our goal was to find cool stars, 4000K $<$ \teff~$<$ 4700K.  Of the
134 stars presented in this paper, 44 had \teff~$\le4700$K and given the
uncertainties in \teff~we note that 69 of the 134 stars had \teff~$\le4800$K.  
Figure \ref{histogram_teff} shows the number of stars versus
\teff~for this study and the \citet{carney94} study.  Our sample includes
cooler stars than the Carney study which highlights the
different temperature regimes of interest.  In addition
to the stars presented here, we have observations
of a further 100 stars which are simply too cool for equivalent width
analysis.  Molecular bands make identification of the continuum
virtually impossible and finding unblended 
Fe {\scshape i} and Fe {\scshape ii} lines is problematic.  
Given the strength of the TiO bands, we assume that 
these stars are all cooler than 4500K.  We are exploring various
techniques for determination of the stellar parameters.  
The strength of the TiO bandheads in some stars suggests
rather low values for \teff, thus we reconfirm Ryan's
finding regarding Luyten's photometry that "the values 
tabulated in the NLTT catalog must be regarded as approximate
only".  

\subsection{Metallicity}

Figure \ref{histogram_fe} shows the number of stars versus metallicity
for this study, the \citet{carney94} sample, and the
\citet{ryan91} sample.  
It is clear that our sample is considerably
smaller than the Ryan and Carney studies.  
The distributions are similar between the
Carney sample and our study which is rather interesting.  Our 
complete list of NLTT candidates comprised 4445 stars
of which we observed some 230 stars.  We derived stellar parameters for 
134 stars and found 27 with [Fe/H] $\le-1.50$, with several 
having prior metallicity estimates.  Presumably, if
we were to observe the remaining $4000+$ candidates, we would find 
$\sim 500$ stars with [Fe/H] $\le-1.50$.  
The distribution of the Ryan sample differs from this study and
from the Carney study.  The Ryan sample peaks at a lower metallicity where this
is simply a selection effect.  The original \citet{ryan89} sample
provided broadband photometry where metal-poor candidates were
identified by the ultraviolet excess.  The stars re-observed
in \citet{ryan91} were those stars identified in \citet{ryan89}
as having an ultraviolet excess corresponding to [Fe/H]$<-1.2$, that is,
$\delta(U-B)_{0.6}>0.2$.  

For the 27 stars we observed with
[Fe/H] $\le-1.50$, there were 11 sufficiently cool to provide
MgH features.  These were the cool subdwarfs with molecular
features that we were interested in finding.  Again, if were to
observe the remaining $4000+$ NLTT candidates there would
presumably be $\sim 200$ such objects.

In our sample, the lack of cool subdwarfs with [Fe/H] $\le -2$
could be attributed to 2 reasons.  Firstly, the volume we are 
sampling may not be large enough and secondly, our ability to
identify genuine subdwarfs may be inefficient.  Regarding the volume we are sampling, 
according to the \citet{girardi2000} isochrones, for stars with 
\teff~= 4500K and [Fe/H] $=-1.68$, our $m_{R} \simeq V = 13$ limit
corresponds to a maximum distance of merely $\sim75$ parsecs.
This volume further decreases when we consider even more metal-poor stars.
More importantly, applying our magnitude limit to a star with 
\teff~= 5500K and [Fe/H] $=-1.68$ corresponds to a maximum distance
of $\sim220$ parsecs.  This represents a volume 27 times larger
than for our cooler targets.  \citet{carney94} considered stars
as faint as V=16 which combined with the intrinsic brightness of warmer targets results
in a considerably larger volume.  Obviously the number of
cool metal-poor stars in a particular volume increases as the volume
increases.  Regarding the efficiency of identifying candidate subdwarfs,
recent work by \citet{salim02} provides more accurate 
$V-J$ photometry for NLTT stars based on
2MASS infrared photometry and USNO-A optical photometry.  Combining 
the precise photometry with
the proper-motions generates a reduced proper-motion diagram with
well defined white dwarf, subdwarf, and disk dwarf 
populations.  From this revised NLTT catalog and proper-motion
diagram we expect that these subdwarf candidates will contain
a considerably higher fraction of genuine subdwarfs than our
original list of candidates.
Observations of candidate subdwarfs selected from 
the \citet{salim02} $V-J$ reduced proper-motion diagram will
be made.

Alternatively, we could consider what volume is needed
to obtain a large number of stars with $-3<$[Fe/H]$<-2$ and
4000K $<$ \teff~$<$ 4500K.  Our observed NLTT candidates provided just 
1 star within the desired range of parameters, G 39-36 (\teff=4200K,
[Fe/H]=$-2.5$).  From the \citet{yi01} isochrones, we estimate that
G 39-36 lies at a distance of 27 parsecs from the sun.  Given the
stellar parameters of G 39-36, our V=13 magnitude limit represents a maximum
volume of merely 36 parsecs.  Since we observed
around 200 of the 4445 NLTT candidates, perhaps there are 10 to 20 cool
subdwarfs amongst our list of NLTT candidates.  If we conservatively estimate 
that there are 5 cool subdwarfs ($-3<$[Fe/H]$<-2$, 4000K $<$ \teff $<$ 4500K)
within a radius of 50 parsecs around the sun, then we expect around 300 
cool subdwarfs within a radius of 200 parsecs.  A search for
subdwarfs similar to G 39-36 out to a distance of 200 parsecs would
require a magnitude limit of V$\simeq17$.  Such a limit implies a
low resolution, low signal-to-noise search technique or a photometric
search to weed out the intruders.

\subsection{Hipparcos}

\begin{figure}
\epsscale{0.9}
\plotone{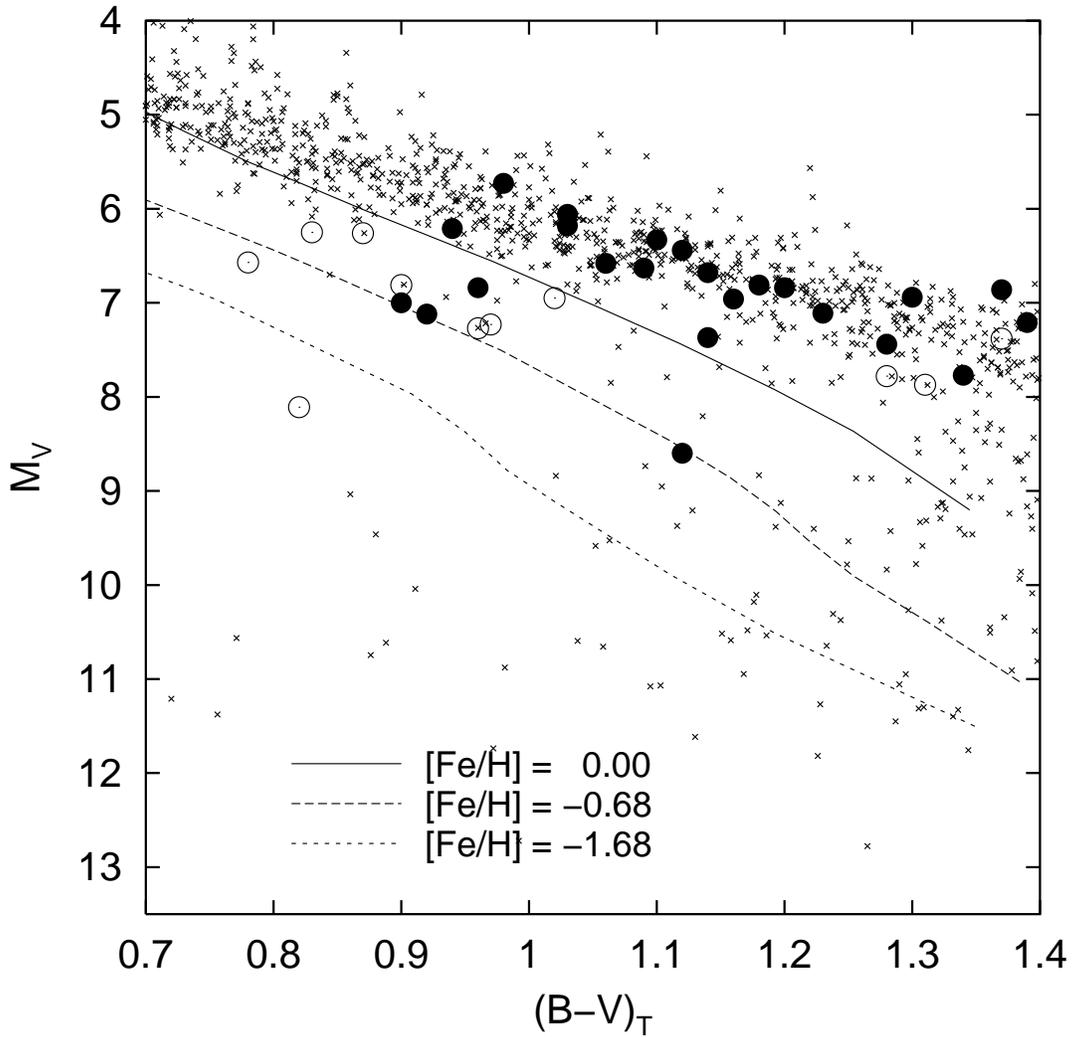}
\caption{The M$_V$, $(B-V)_T$ color-magnitude diagram with crosses marking \hipparcos
stars with $\sigma_\pi/\pi<15$ and $\pi>32$mas.  Closed circles represent observed stars 
with [Fe/H]$\ge-1.0$ while open circles represent [Fe/H]$<-1.0$.  
\citet{girardi2000} isochrones using Johnson $(B-V)$
for [Fe/H]=$0.0,-0.68,-1.68$ are overplotted.  $(B-V)_T$ is 
increasingly redder than the
Johnson $(B-V)$ as the stars get cooler as noted 
by \citet{reid01}. \label{hipparcos}}
\end{figure}

Having observed a number of cool stars contained in the \hipparcos catalog,
we can comment upon the results of \citet{reid01}.  The majority
of \hipparcos stars were selected by their large reduced proper-motion from the NLTT
catalog rather than the position in the color-magnitude diagram.
\citet{reid01} claimed that in a color-magnitude diagram produced
from \hipparcos parallaxes, the majority of stars with ($B-V$) $\le0.8$ situated below
the solar metallicity main sequence were incorrectly positioned
due to errors in the colors.  These stars, located where subdwarfs ought to lie,
were metal rich disk dwarfs rather than subdwarfs.
\citet{reid01} also showed that
the Tycho colors ($B-V$)$_T$ are redder than the Johnson ($B-V$) colors.
In Figure \ref{hipparcos}, we take the Tycho colors and \hipparcos
parallaxes to produce a color-magnitude diagram overlaying the
stars we observed.  Nearby stars ($\pi > 32$ mas) with accurate
parallaxes ($\sigma_\pi/\pi<0.15$) are included to represent the
disk population, [Fe/H]$\simeq0$.  Our few observations confirm
and extend the \citet{reid01} findings in the regime $0.8 < B-V < 1.4$.
Metal-poor stars with [Fe/H] $< -1.0$ are located amongst disk stars.  Also,
metal-rich stars with [Fe/H] $>-1.0$ are offset from the disk main sequence.  The effect
of the redder Tycho colors is evident as the disk main sequence is systematically
offset from the \citet{girardi2000} [Fe/H]=0 isochrone.
We intend to observe additional \hipparcos
subdwarf candidates in the range 0.8 $<$ ($B-V$)$_T$ $<$ 1.2 not only to
find cool subdwarfs but to further test the findings made by \citet{reid01}.

\section{Concluding remarks}
\label{sec:remarks}

We present stellar parameters for 134 candidate subdwarfs selected by
their reduced proper-motion, offset from the solar metallicity,
or from the literature.  Our goal was to provide a large database of
cool subdwarfs.  Our selection criteria were successful in identifying
cool stars (69 with \teff$<4800$K) and metal-poor stars (27 with
[Fe/H]$\le-1.50$).  Of our sample, 11 stars were sufficiently
cool to provide measurable MgH lines with [Fe/H]$<-1.50$.
Armed with a sample of cool subdwarfs, we can begin to exploit their unique 
qualities.  For a subset 
of our sample, we have measured the Mg
isotopic abundance ratios and compared the observed trends
with predictions from models of Galactic chemical evolution.  This work will be
presented in a future paper.  We also intend to make further observations
of candidate subdwarfs.  Targets will be selected from the revised
NLTT catalog \citep{salim02}.  Cool \hipparcos subdwarf candidates
will be observed and we anticipate that Christlieb and collaborators
will identify cool subdwarfs in the Hamburg/ESO survey data.

\acknowledgments

We thank George Preston for helpful comments and the staff of 
the Hobby-Eberly telescope for service observations of our targets.
We acknowledge support via the grant F-634 from the Robert A. Welch
Foundation of Houston, Texas.
This research has made use of the SIMBAD database,
operated at CDS, Strasbourg, France and
NASA's Astrophysics Data System.

\clearpage

\clearpage

\clearpage

\clearpage

\clearpage

\clearpage

\clearpage

\clearpage

\clearpage

\clearpage

\clearpage

\input{tab1}

\input{tab2}
\input{tab3}

\end{document}

%% file: tab1.tex
\begin{deluxetable}{lcrccrccrrrrrrrc} 
\tabletypesize{\tiny}
\rotate
\tablecolumns{15} 
\tablewidth{0pc} 
\tablecaption{Basic data and derived parameters for objects \label{param}}
\tablehead{ 
\colhead{Name}          & \colhead{Ra\tablenotemark{a}}     &
\colhead{Dec\tablenotemark{b}}           & 
\colhead{Resolving}     & \colhead{\teff} &
\colhead{[Fe/H]}        & \colhead{log g} &
\colhead{$\xi_t$}       & \colhead{U$_{\rm LSR}$} &
\colhead{$\sigma_U$}    & \colhead{V$_{\rm LSR}$} &
\colhead{$\sigma_V$}    & \colhead{W$_{\rm LSR}$} &
\colhead{$\sigma_W$}    & \colhead{Luyten color}  &
\colhead{Molecular}		\\ 
\cline{8-14} \\
\colhead{}              & \colhead{2000}     &
\colhead{2000}          & 
\colhead{Power\tablenotemark{c}}         & \colhead{(K)} &
\colhead{}              & \colhead{(cm s$^{-2}$)} &
\multicolumn{7}{c}{(km/s)}  & \colhead{or source} &
\colhead{Features}
}
\startdata 
PLX 5805 & 000155 & 260015 & 30000 & 4600 & -1.72 & 4.25 & 1.0 & 120 & 10 & -97 & 3 & -26 & 8 & Yale & Y \\
LP 524-64 & 000800 & 081642 & 30000 & 4600 & -1.04 & 5.00 & 1.0 & -90 & 30 & -95 & 25 & 8 & 13 & k-m & Y \\
BD -14 142 & 004822 & -140713 & 60000 & 6200 & -1.21 & 4.00 & 1.0 & 50 & 30 & -225 & 110 & 2 & 25 & g-k & N \\
LP 646-24 & 005006 & -042600 & 30000 & 4800 & -1.01 & 4.50 & 0.7 & -16 & 5 & -120 & 45 & -38 & 16 & g-k & Y \\
G 72-34 & 014604 & 355448 & 11000 & 4800 & -2.23 & 4.75 & 0.9 & 45 & 10 & -265 & 165 & -243 & 230 & g-k & N \\
BD -4 290 & 015218 & -032643 & 60000 & 4900 & -0.01 & 4.50 & 0.7 & 50 & 10 & -2 & 1 & -10 & 5 & k & Y \\
BD +8 335 & 021120 & 093724 & 60000 & 5100 & -0.98 & 4.50 & 0.7 & -35 & 2 & -65 & 13 & -54 & 5 & g-k & Y \\
HIP 10412 & 021411 & 013312 & 30000 & 4600 & -0.40 & 4.75 & 1.2 & -65 & 6 & -24 & 5 & 2 & 3 & Yale & Y \\
Cool 340 & 021726 & 270827 & 60000 & 4800 & -0.40 & 5.00 & 0.8 & 14 & 6 & -98 & 17 & 16 & 7 & g-k & Y \\
BD +4 415 & 023435 & 052647 & 60000 & 4750 & -0.63 & 4.50 & 0.6 & 122 & 6 & -52 & 3 & -15 & 6 & k & Y \\
G 78-26 & 031627 & 380556 & 30000 & 4200 & -1.01 & 4.75 & 0.6 & 65 & 5 & -245 & 15 & -17 & 6 & Gizis97 & Y \\
GJ 1064A & 034702 & 412538 & 60000 & 5000 & -1.15 & 4.50 & 0.6 & -106 & 3 & -109 & 7 & -71 & 4 & FJ98 & Y \\
GJ 1064B & 034703 & 412542 & 60000 & 4800 & -1.15 & 4.50 & 0.6 & -105 & 6 & -130 & 17 & -80 & 9 & FJ98 & Y \\
G 175-7 & 034828 & 511406 & 11000 & 5000 & -1.76 & 4.00 & 1.0 & -164 & 240 & -515 & 400 & -23 & 30 & g-k & N \\
BD -4 680 & 035154 & -034911 & 11000 & 5800 & -1.89 & 4.00 & 1.0 & -232 & 40 & -185 & 58 & 77 & 58 & g-k & N \\
GL 158 & 040315 & 351624 & 60000 & 4800 & -1.79 & 5.00 & 1.0 & -50 & 20 & -185 & 4 & 26 & 1 & FJ98 & Y \\
G 39-36 & 044807 & 330936 & 60000 & 4200 & -2.50 & 4.00 & 1.0 & -14 & 8 & -95 & 25 & 82 & 20 & g-k & Y \\
G 81-41 & 045505 & 454405 & 60000 & 4550 & -0.96 & 4.50 & 0.6 & -8 & 4 & -79 & 12 & 72 & 11 & k & Y \\
BD -10 1085 & 050422 & -100859 & 60000 & 5100 & -0.50 & 4.50 & 0.8 & -6 & 3 & -70 & 4 & -27 & 1 & g-k & Y \\
BD +19 869 & 051253 & 194320 & 60000 & 4600 & -1.01 & 4.50 & 0.5 & 0 & 1 & -110 & 9 & -5 & 1 & k & Y \\
PLX 1219 & 052310 & 331130 & 30000 & 4600 & -1.79 & 4.00 & 0.6 & -240 & 15 & -125 & 90 & 9 & 5 & Yale & Y \\
HIP 27928 & 055434 & -092334 & 60000 & 4200 & -0.93 & 4.00 & 0.6 & -70 & 3 & 25 & 3 & 45 & 4 & Yale, HIP & Y \\
G 103-27 & 061648 & 295706 & 60000 & 5600 & -0.95 & 4.50 & 0.6 & 0 & 5 & -145 & 95 & -118 & 80 & K & N \\
G 101-34 & 062004 & 382044 & 60000 & 5000 & -1.50 & 2.75 & 1.0 & -336 & 374 & -1425 & 4930 & 109 & 206 & g\tablenotemark{d} & N \\
HIP 30567 & 062530 & 484340 & 60000 & 4700 & -0.60 & 4.75 & 0.6 & 8 & 4 & -104 & 12 & -46 & 5 & k & Y \\
G 103-50 & 064008 & 282712 & 60000 & 4700 & -2.20 & 4.50 & 0.4 & -147 & 38 & -71 & 3 & 15 & 24 & AAM96, CLLA94 & Y \\
AC +78 2199 & 065249 & 781230 & 60000 & 5000 & -0.03 & 4.50 & 0.8 & -120 & 19 & -4 & 4 & -114 & 23 & g-k & Y \\
G 88-1 & 065828 & 185949 & 60000 & 4600 & -0.84 & 5.00 & 1.0 & 2 & 3 & -45 & 6 & 41 & 4 & g-k & Y \\
BD -17 1716 & 065929 & -171517 & 60000 & 4600 & -0.08 & 4.50 & 0.4 & -53 & 3 & 44 & 3 & -25 & 3 & k & Y \\
G 87-27 & 071008 & 371634 & 60000 & 5100 & -0.61 & 3.50 & 0.9 & -336 & 374 & -1425 & 4930 & 109 & 206 & CLLA94 & Y \\
G 87-35 & 072044 & 292048 & 30000 & 5300 & -0.53 & 4.50 & 0.8 & -51 & 5 & -123 & 21 & 50 & 6 & g-k & Y \\
G 193-42 & 072222 & 503318 & 11000 & 5000 & -0.94 & 4.25 & 1.0 & -155 & 35 & -275 & 140 & 37 & 10 & g-k & N \\
G 193-49 & 072722 & 522612 & 30000 & 5000 & -0.39 & 5.00 & 0.9 & -72 & 26 & 35 & 14 & -118 & 58 & k & Y \\
G 90-4 & 073023 & 271618 & 11000 & 5600 & -0.81 & 4.50 & 1.0 & 7 & 10 & -150 & 120 & -108 & 80 & g-k & N \\
GSC 03790-02030 & 074534 & 565722 & 30000 & 5600 & -1.16 & 4.00 & 0.7 & -43 & 37 & -84 & 50 & -59 & 32 & K & N \\
LP 208-60 & 082144 & 384713 & 11000 & 5800 & -0.22 & 4.00 & 1.0 & 50 & 35 & -280 & 255 & 34 & 37 & g-k & N \\
G 115-1 & 082217 & 410424 & 30000 & 4800 & -1.33 & 4.50 & 0.6 & -75 & 3 & -84 & 23 & 34 & 4 & g-k & Y \\
LP 545-17 & 082540 & 051528 & 11000 & 5200 & -1.01 & 4.75 & 1.0 & 52 & 105 & -95 & 50 & 177 & 120 & k-m & N \\
PLX 2019 & 082941 & 014448 & 30000 & 4600 & -1.72 & 4.50 & 1.0 & 237 & 101 & -334 & 104 & -11 & 14 & Yale & Y \\
PLX 2023.01 & 083000 & -095401 & 30000 & 4400 & -0.39 & 3.90 & 1.0 & -12 & 1 & 4 & 1 & 10 & 1 & Yale & Y \\
LP 209-14 & 083110 & 421348 & 11000 & 5400 & -1.36 & 4.75 & 1.0 & -155 & 16 & -270 & 265 & 92 & 15 & K & N \\
G 113-49 & 083512 & 032800 & 11000 & 5100 & -0.83 & 4.00 & 1.0 & 95 & 120 & -180 & 110 & 77 & 35 & k-m & N \\
G 9-31 & 085245 & 223330 & 11000 & 5600 & -1.00 & 4.25 & 1.0 & -96 & 16 & -150 & 45 & -93 & 45 & g-k & N \\
LP 36-78 & 085629 & 705814 & 60000 & 5150 & 0.25 & 4.50 & 0.7 & -87 & 16 & -56 & 23 & 31 & 2 & k & Y \\
LP 786-61 & 090514 & -183206 & 60000 & 5800 & -0.76 & 4.00 & 0.8 & 63 & 23 & 16 & 3 & 47 & 20 & g-k & N \\
ROSS 94 & 094722 & 261813 & 60000 & 4600 & 0.00 & 3.50 & 1.0 & -24 & 2 & -49 & 8 & -18 & 4 & k & Y \\
G 116-55 & 094819 & 340712 & 30000 & 5600 & -2.10 & 4.00 & 1.0 & -190 & 90 & -165 & 130 & -38 & 90 & g-k & N \\
G 43-7 & 095013 & 050905 & 30000 & 4950 & -1.02 & 4.50 & 0.8 & 113 & 19 & -41 & 11 & 12 & 5 & g-k & Y \\
G 161-84 & 095139 & -035000 & 60000 & 4700 & -1.31 & 4.75 & 0.4 & -199 & 53 & -96 & 17 & 131 & 15 & CLLA94 & Y \\
LP 788-55 & 095433 & -192100 & 60000 & 4700 & -0.53 & 4.50 & 0.7 & 66 & 18 & -4 & 6 & -5 & 4 & g-k & Y \\
WOLF 334 & 095744 & 323654 & 11000 & 5200 & -1.71 & 4.50 & 1.0 & -450 & 320 & -375 & 335 & -163 & 300 & g-k & N \\
LP 315-12 & 095746 & 264522 & 60000 & 5100 & -0.98 & 4.50 & 0.6 & -122 & 15 & -12 & 4 & -14 & 11 & g-k & Y \\
LP 429-17 & 100108 & 141842 & 60000 & 4750 & -0.10 & 4.50 & 0.4 & 47 & 10 & -47 & 13 & -15 & 3 & g-k & Y \\
BD +53 1395 & 101357 & 523024 & 60000 & 4500 & -1.04 & 4.50 & 0.5 & 16 & 1 & -79 & 3 & 12 & 1 & k & Y \\
BD +12 2201 & 102220 & 120845 & 60000 & 4500 & -0.23 & 4.50 & 0.9 & -57 & 3 & -17 & 1 & -9 & 2 & k & Y \\
LP 790-19 & 102607 & -175843 & 60000 & 4300 & -0.51 & 5.00 & 0.6 & -111 & 7 & -46 & 1 & 9 & 2 & k & Y \\
BD -9 3104 & 103529 & -102236 & 30000 & 4900 & -0.60 & 4.50 & 0.8 & 130 & 32 & -82 & 14 & -9 & 9 & k & Y \\
G 119-21 & 103725 & 285548 & 30000 & 4600 & -0.48 & 4.50 & 0.6 & -81 & 24 & 58 & 15 & -42 & 10 & k & Y \\
BD +31 2175 & 104016 & 304855 & 60000 & 4800 & -0.17 & 4.75 & 1.1 & -90 & 4 & -35 & 2 & -13 & 3 & k & Y \\
PLX 2529.1 & 105203 & -000938 & 60000 & 4800 & -0.58 & 4.50 & 0.5 & 39 & 3 & -6 & 3 & -52 & 2 & Yale, HIP & Y \\
G 119-45 & 105721 & 305948 & 30000 & 4800 & 0.14 & 4.50 & 0.7 & -177 & 39 & -85 & 22 & -53 & 20 & k & Y \\
G 58-36 & 110313 & 220024 & 30000 & 4600 & -0.05 & 4.50 & 0.8 & -115 & 35 & -47 & 15 & 15 & 16 & k & Y \\
G 253-46 & 110806 & 822454 & 60000 & 5200 & -0.57 & 4.50 & 0.7 & 0 & 2 & -35 & 6 & 56 & 7 & g-k & Y \\
BD -10 3216 & 111111 & -105703 & 60000 & 4500 & -1.19 & 4.50 & 0.4 & -122 & 3 & -16 & 1 & 34 & 1 & k-m & Y \\
HIP 55128 & 111712 & 172927 & 30000 & 4900 & -0.38 & 4.50 & 0.9 & -124 & 26 & -55 & 14 & -53 & 12 & k & Y \\
LP 792-12 & 111858 & -171548 & 60000 & 4800 & -0.39 & 5.00 & 0.6 & -77 & 14 & -34 & 2 & 35 & 2 & g-k & Y \\
BD +15 2325 & 112219 & 142644 & 60000 & 4800 & -0.58 & 4.50 & 0.6 & -125 & 12 & -46 & 5 & -9 & 5 & k & Y \\
G 122-22 & 112400 & 453234 & 30000 & 4400 & -1.36 & 4.50 & 0.9 & 63 & 15 & -36 & 8 & 58 & 8 & k & Y \\
G 163-80 & 112507 & -055624 & 30000 & 4700 & 0.05 & 4.75 & 0.6 & -114 & 23 & -55 & 9 & 0 & 8 & k & Y \\
ROSS 109 & 112715 & 593318 & 60000 & 5300 & -1.53 & 4.00 & 0.6 & -350 & 300 & -645 & 700 & 217 & 140 & K & N \\
LP 733-14 & 113829 & -135006 & 60000 & 4700 & -0.57 & 4.75 & 0.7 & -15 & 3 & -35 & 11 & -41 & 10 & g-k & Y \\
LP 734-54 & 115754 & -094848 & 60000 & 4800 & -0.66 & 4.50 & 0.4 & -78 & 13 & -15 & 4 & 5 & 2 & g-k & Y \\
WOLF 1424 & 120019 & 203543 & 60000 & 4600 & -1.19 & 4.50 & 0.4 & -28 & 4 & -120 & 25 & -33 & 6 & k & Y \\
LP 734-101 & 121655 & -114906 & 60000 & 5500 & 0.06 & 4.50 & 0.7 & -86 & 16 & -36 & 10 & -9 & 4 & g-k & Y \\
G 13-29 & 121920 & 022642 & 30000 & 4800 & -0.66 & 4.50 & 0.6 & -6 & 4 & -101 & 30 & -8 & 15 & g-k & Y \\
G 198-61 & 123523 & 374430 & 30000 & 5300 & -0.12 & 4.50 & 1.0 & 176 & 36 & -42 & 11 & 21 & 7 & k & Y \\
G 149-9 & 124803 & 274548 & 30000 & 4900 & -0.62 & 4.75 & 0.8 & 24 & 16 & -315 & 150 & -86 & 1 & k & Y \\
HIP 62627 & 124956 & 711139 & 60000 & 4800 & -0.24 & 5.00 & 1.2 & -8 & 2 & -93 & 3 & -17 & 2 & HIP & Y \\
G 149-18 & 125246 & 222700 & 30000 & 5000 & -0.03 & 4.50 & 0.8 & -214 & 44 & -33 & 11 & 29 & 2 & k & Y \\
W 453 & 125749 & 054554 & 30000 & 4600 & -0.04 & 4.75 & 0.9 & -71 & 17 & -38 & 13 & 2 & 2 & k & Y \\
HD 114095 & 130826 & -071830 & 60000 & 4730 & -0.71 & 2.40 & 1.3 & 65 & 43 & -290 & 166 & -49 & 32 & AAM96 & Y \\
BD +68 714 & 131021 & 672941 & 30000 & 4900 & -0.05 & 4.50 & 1.0 & -96 & 3 & -70 & 2 & 7 & 1 & K1 & Y \\
ROSS 466 & 131514 & -110112 & 30000 & 5000 & -0.86 & 4.75 & 0.9 & -71 & 17 & -115 & 65 & -103 & 30 & g-k & Y \\
LHS 2715 & 131857 & -030418 & 60000 & 4400 & -1.56 & 4.00 & 0.6 & -40 & 7 & -120 & 7 & 115 & 1 & Gizis97 & Y \\
LP 172-89 & 132638 & 464851 & 30000 & 4800 & -0.49 & 4.50 & 0.6 & -124 & 28 & -34 & 9 & 0 & 2 & g-k & Y \\
LP 738-45 & 133814 & -154714 & 30000 & 4400 & -0.12 & 4.50 & 0.8 & -81 & 17 & -28 & 14 & 14 & 4 & k & Y \\
G 255-44 & 135536 & 740012 & 60000 & 4900 & -1.30 & 4.50 & 0.7 & -107 & 17 & -25 & 15 & 74 & 7 & k & Y \\
G 65-22 & 140144 & 085517 & 60000 & 5000 & -1.66 & 4.50 & 0.4 & 10 & 8 & -225 & 76 & -83 & 25 & AAM96, CLLA94 & Y \\
BD +30 2490 & 141124 & 300502 & 30000 & 4600 & -0.18 & 4.75 & 0.6 & -68 & 5 & -88 & 8 & 35 & 3 & k & Y \\
G 239-12 & 141853 & 731412 & 60000 & 5800 & -2.62 & 4.00 & 1.0 & 74 & 25 & -333 & 160 & 127 & 170 & K & N \\
HIP 70152 & 142114 & 085816 & 60000 & 4800 & -0.05 & 5.00 & 1.0 & 50 & 7 & -36 & 5 & -20 & 4 & RWMRKM01 & Y \\
HIP 70715 & 142745 & 235027 & 60000 & 4400 & -0.11 & 4.50 & 0.6 & -110 & 9 & -66 & 6 & -9 & 4 & HIP & Y \\
LP 500-92 & 143413 & 123457 & 30000 & 4500 & -0.38 & 4.50 & 0.9 & -134 & 9 & -37 & 4 & 10 & 5 & k-m & Y \\
LP 175-8 & 143610 & 450859 & 30000 & 4900 & -0.31 & 4.75 & 0.6 & -123 & 25 & -65 & 14 & 33 & 10 & k & Y \\
HD 131287 & 145004 & 603936 & 30000 & 4800 & -0.04 & 4.50 & 1.0 & -4 & 1 & -12 & 1 & -13 & 1 & k & Y \\
BD +23 2751 & 145342 & 232043 & 60000 & 4700 & -0.51 & 4.50 & 0.6 & -76 & 2 & -62 & 2 & 20 & 1 & k & Y \\
LP 502-8 & 150914 & 143123 & 30000 & 5000 & -0.95 & 4.50 & 0.6 & -80 & 17 & -119 & 25 & 113 & 16 & g-k & Y \\
HIP 74235 & 151013 & -162246 & 60000 & 4900 & -1.51 & 4.50 & 0.8 & 295 & 2 & -505 & 18 & -62 & 10 & HIP & Y \\
LP 802-56 & 151912 & -210046 & 60000 & 5200 & -0.88 & 4.50 & 0.6 & -143 & 7 & -130 & 32 & -82 & 7 & k & Y \\
BD +2 2944 & 151919 & 014555 & 30000 & 6000 & -0.02 & 3.50 & 1.8 & 81 & 1 & -8 & 1 & -3 & 1 & K4 & N \\
ROSS 804 & 154529 & -134918 & 30000 & 4600 & -0.93 & 4.50 & 0.6 & 120 & 85 & -285 & 360 & -108 & 200 & k & Y \\
G 225-50 & 161445 & 552549 & 30000 & 4600 & -0.13 & 4.50 & 1.0 & -60 & 13 & -43 & 6 & 8 & 7 & g-k & Y \\
LP 330-7 & 162306 & 320835 & 30000 & 4800 & -0.22 & 4.50 & 0.6 & 74 & 16 & 4 & 6 & 10 & 6 & g-k & Y \\
G 202-57 & 163337 & 511330 & 30000 & 4900 & -1.05 & 4.50 & 0.6 & 180 & 82 & -65 & 32 & 49 & 18 & g-k & Y \\
G 17-25 & 163442 & -041345 & 60000 & 4950 & -1.46 & 4.50 & 0.3 & 11 & 11 & -115 & 24 & -29 & 6 & CLLA94 & Y \\
BD -14 4454 & 163614 & -151012 & 30000 & 4600 & -0.95 & 4.50 & 0.6 & -111 & 2 & -90 & 8 & 35 & 5 & k & Y \\
LP 445-55 & 164258 & 192209 & 30000 & 5200 & -1.80 & 4.00 & 0.6 & 118 & 24 & 38 & 8 & 84 & 13 & g-k & N \\
LP 447-2 & 170805 & 175738 & 30000 & 5800 & -1.79 & 4.00 & 0.9 & -146 & 7 & -272 & 26 & 46 & 18 & g-k & N \\
LP 747-18 & 171644 & -130112 & 30000 & 5800 & -0.95 & 4.50 & 0.8 & 71 & 51 & -322 & 327 & 41 & 29 & g-k & N \\
LP 447-34 & 171708 & 180706 & 30000 & 4800 & -0.35 & 4.50 & 0.9 & 53 & 18 & -45 & 20 & 11 & 4 & g-k & Y \\
G 19-25 & 172559 & -024436 & 60000 & 4900 & -2.01 & 4.50 & 0.4 & 116 & 86 & -93 & 85 & 9 & 6 & AAM96, CLLA94 & Y \\
LP 808-11 & 173353 & -165854 & 30000 & 5500 & -1.23 & 4.00 & 0.7 & -35 & 40 & -225 & 200 & -78 & 65 & g-k & N \\
BD -8 4501 & 174728 & -084648 & 30000 & 5800 & -1.80 & 4.00 & 0.8 & 126 & 12 & -45 & 16 & -156 & 38 & g-k & N \\
G 204-47 & 180929 & 471324 & 30000 & 5200 & -0.88 & 4.50 & 1.0 & 170 & 126 & -137 & 65 & 73 & 59 & K & N \\
BD +5 3640 & 181222 & 052404 & 60000 & 4950 & -1.34 & 4.50 & 0.4 & 61 & 13 & -216 & 21 & 38 & 5 & AAM96 & Y \\
BD -17 5287 & 184057 & -170954 & 30000 & 5800 & -0.70 & 4.50 & 1.0 & -60 & 25 & -155 & 80 & 77 & 35 & K & N \\
G 227-44 & 184143 & 583430 & 30000 & 5400 & -1.45 & 4.50 & 1.0 & -314 & 250 & -210 & 11 & -37 & 42 & K & N \\
G 207-23 & 191613 & 370420 & 30000 & 5000 & -1.90 & 4.00 & 0.6 & -175 & 6 & -319 & 1 & 41 & 15 & K & N \\
HD 181007 & 191928 & -202540 & 60000 & 4700 & -1.93 & 1.50 & 1.6 & 65 & 43 & -290 & 166 & -49 & 32 & None\tablenotemark{e} & N \\
LP 752-18 & 192547 & -110954 & 30000 & 5800 & -1.00 & 4.00 & 1.0 & 60 & 25 & -75 & 40 & 17 & 10 & g-k & N \\
LP 753-29 & 195004 & -131912 & 30000 & 5400 & -1.88 & 4.00 & 1.0 & -65 & 26 & -155 & 76 & 77 & 35 & g-k & N \\
G 23-14 & 195150 & 053646 & 60000 & 5000 & -1.51 & 3.00 & 1.0 & 116 & 86 & -93 & 85 & 9 & 6 & g\tablenotemark{f} & N \\
LP 575-39 & 202326 & 055050 & 30000 & 4700 & -0.42 & 4.50 & 0.6 & -90 & 13 & -11 & 7 & -19 & 12 & g-k & Y \\
LP 575-40 & 202356 & 052444 & 30000 & 4800 & -0.10 & 4.50 & 1.0 & -110 & 25 & 20 & 15 & -29 & 18 & k-m & Y \\
LP 815-43 & 203814 & -202554 & 30000 & 6400 & -2.70 & 4.00 & 1.0 & 140 & 140 & -290 & 265 & 36 & 27 & g-k & N \\
HD 200968 & 210710 & -135523 & 30000 & 5000 & -0.16 & 4.50 & 1.2 & -52 & 1 & -14 & 1 & 2 & 1 & KF99 & Y \\
BD +22 4567 & 221031 & 224749 & 30000 & 4800 & -0.33 & 4.75 & 1.1 & 75 & 4 & -5 & 1 & 70 & 3 & k & Y \\
BD +30 4633 & 221206 & 313341 & 60000 & 4600 & -0.01 & 4.50 & 0.8 & 92 & 6 & -35 & 1 & -6 & 2 & k-m & Y \\
HIP 109801 & 221424 & -084442 & 30000 & 4600 & -1.70 & 4.50 & 1.0 & 130 & 50 & -225 & 75 & -43 & 25 & HIP & Y \\
ROSS 237 & 225330 & 274518 & 30000 & 4900 & -1.49 & 4.25 & 0.9 & -118 & 25 & -315 & 16 & 8 & 25 & k & Y \\
ROSS 242 & 230849 & 270054 & 60000 & 4700 & -0.94 & 4.50 & 0.4 & 50 & 14 & -130 & 16 & -67 & 30 & k & Y \\
LP 702-79 & 232332 & -060012 & 30000 & 4800 & 0.05 & 4.50 & 1.1 & -150 & 60 & -70 & 25 & 1 & 22 & g-k & Y \\
HIP 115664 & 232556 & 291141 & 30000 & 4600 & -0.90 & 5.00 & 0.6 & -85 & 16 & -215 & 20 & -78 & 25 & HIP & Y \\
BD +28 4634 & 234510 & 293343 & 60000 & 5000 & -0.30 & 4.50 & 0.7 & -138 & 3 & 6 & 2 & -60 & 1 & k & Y \\
\enddata 

\tablenotetext{a}{hhmmss}
\tablenotetext{b}{ddmmss}
\tablenotetext{c}{HET data taken with R=11000, 2.7m data taken with R=30000,60000}
\tablenotetext{d}{Star chosen from \citet{carney94} and \citet{alonso96} but also in NLTT catalogue}
\tablenotetext{e}{Star chosen from \citet{alonso96} not in NLTT catalogue}
\tablenotetext{f}{Star chosen from \citet{carney94} but also in NLTT catalogue}

\tablerefs{
(AAM96) = \citet{alonso96};
(CLLA94) = \citet{carney94};
(FJ98) = \citet{fj98};
(KF99) = Kotoneva \& Flynn (1999, private communication)
(Gizis97) = \citet{gizis97};
(HIP) = \hipparcos color-magnitude diagram;
(RWMRKM01) = \citet{reid01};
(Yale) = Yale color-magnitude diagram
}

\end{deluxetable}

%% file: tab2.tex
\begin{deluxetable}{lcccccr} 
\tabletypesize{\tiny}
\tablecolumns{7} 
\tablewidth{0pc} 
\tablecaption{Comparison with literature\label{comp}}
\tablehead{ 
\colhead{}& \multicolumn{2}{c}{This study} &
\colhead{} & \multicolumn{2}{c}{Literature} &
\colhead{}\\
\cline{2-3}  \cline{5-6}\\
\colhead{Star}& \colhead{\teff} &
\colhead{[Fe/H]}& \colhead{} &
\colhead{\teff} & \colhead{[Fe/H]} & 
\colhead{Source}
}
\startdata 
PLX 5805  & 4600 & -1.72 &  &  \nodata  & -1.28 &  SPC93 \\
LP 646-24  & 4800 & -1.01 &  &  \nodata  & -0.24 &  RN91 \\
G 72-34     & 4800 & -2.23 &  & 4689 & -2.21 &  CLLA94 \\
BD +8 335  & 5100 & -0.98 &  & 5224 & -0.80 &  CLLA94 \\
        & 5100 & -0.98 &  &  \nodata  & -0.61 &  Eggen98 \\
BD +4 415  & 4750 & -0.63 &  & 4756 & -0.41 &  AAM96 \\
        & 4750 & -0.63 &  &  \nodata  & -0.43 &  Reid01 \\
        & 4750 & -0.63 &  &  \nodata  & -0.39 &  Eggen98 \\
G 78-26  & 4200 & -1.01 &  &  \nodata  & -1.20 &  Gizis97 \\
GJ 1064A  & 5000 & -1.15 &  & 5049 & -1.02 &  CLLA94 \\
        & 5000 & -1.15 &  & 5200 & -1.05 &  TL99 \\
        & 5000 & -1.15 &  &  \nodata  & -0.85 &  TI99 \\
        & 5000 & -1.15 &  & 5050 & -0.80 &  Fulbright00 \\
GJ 1064B  & 4800 & -1.15 &  & 4950 & -1.11 &  TL99 \\
        & 4800 & -1.15 &  & 4816 & -1.05 &  CLLA94 \\
        & 4800 & -1.15 &  &  \nodata  & -0.86 &  TI99 \\
BD -4 680   & 5800 & -1.89 &  & 5866 & -2.07 &  AAM96   \\
GL 158  & 4800 & -1.79 &  & 4762 & -1.73 &  CLLA94 \\
        & 4800 & -1.79 &  & 4700 & -1.70 &  Fulbright00 \\
        & 4800 & -1.79 &  &  \nodata  & -1.67 &  TI99 \\
        & 4800 & -1.79 &  & 4775 & -1.81 &  TL99 \\
BD +19 869  & 4600 & -1.01 &  &  \nodata  & -0.55 &  Eggen87 \\
        & 4600 & -1.01 &  &  \nodata  & -0.60 &  Norris86 \\
PLX 1219  & 4600 & -1.79 &  & 4935 & -1.02 &  CLLA94 \\
        & 4600 & -1.79 &  &  \nodata  & -2.00 &  Morrison01 \\
        & 4600 & -1.79 &  & 4739 & -1.61 &  AAM96 \\
G 103-27    & 5600 & -0.95 &  & 5449 & -0.93 &  SPC93  \\
G 101-34    & 5000 & -1.50 &  & 5289 & -1.83 &  AAM96\tablenotemark{a}  \\
        & 5000 & -1.50 &  & 4955 & -1.88 &  CLLA94 \\
G 103-50  & 4700 & -2.20 &  & 4642 & -2.00 &  CLLA94 \\
        & 4700 & -2.20 &  & 4713 & -3.00 &  AAM96 \\
G 88-1  & 4600 & -0.84 &  &  \nodata  & -0.13 &  RN91 \\
        & 4600 & -0.84 &  &  \nodata  & -0.40 &  SPC93 \\
G 87-27  & 5100 & -0.61 &  & 4905 & -1.45 &  CLLA94\tablenotemark{a} \\
        & 5100 & -0.61 &  & 5090 & -0.42 &  GL00 \\
G 87-35  & 5300 & -0.53 &  & 5222 & -0.48 &  CLLA94 \\
        & 5300 & -0.53 &  &  \nodata  & -0.55 &  Eggen98 \\
G 90-4      & 5600 & -0.81 &  & 5513 & -0.86 &  CLLA94 \\
PLX 2019  & 4600 & -1.72 &  & 4740 & -1.13 &  AAM96 \\
G 113-49    & 5100 & -0.83 &  & 5194 & -0.78 &  CLLA94 \\
G 9-31      & 5600 & -1.00 &  & 5546 & -0.94 &  CLLA94 \\
        & 5600 & -1.00 &  & 5569 & -0.85 &  AAM96 \\
G 43-7  & 4950 & -1.02 &  & 5009 & -0.68 &  CLLA94 \\
        & 4950 & -1.02 &  &  \nodata  & -0.68 &  Reid01 \\
G 161-84  & 4700 & -1.31 &  & 4650 & -1.71 &  CLLA94 \\
        & 4700 & -1.31 &  &  \nodata  & -1.57 &  Beers99 \\
WOLF 334    & 5200 & -1.71 &  & 5084 & -1.82 &  CLLA94 \\
LP 792-12  & 4800 & -0.39 &  &  \nodata  & 0.18 &  RN91 \\
G 13-29  & 4800 & -0.66 &  &  \nodata  & 0.04 &  RN91 \\
HD 114095  & 4730 & -0.71 &  & 4650 & -0.60 &  Fulbright00 \\
        & 4730 & -0.71 &  & 4541 & -1.58 &  AAM96\tablenotemark{a} \\
        & 4730 & -0.71 &  & 4680 & -0.35 &  Clementini99 \\
BD +68 714  & 4900 & -0.05 &  &  \nodata  & -0.17 &  Eggen98 \\
LHS 2715  & 4400 & -1.56 &  & 4480 & -0.22 &  AAM96 \\
        & 4400 & -1.56 &  &  \nodata  & -1.20 &  Gizis97 \\
G 65-22  & 5000 & -1.66 &  & 4971 & -1.75 &  AAM96 \\
        & 5000 & -1.66 &  & 4982 & -1.72 &  CLLA94 \\
G 239-12    & 5800 & -2.62 &  & 6025 & -2.56 &  CLLA94 \\
HIP 70152  & 4800 & -0.05 &  &  \nodata  & 0.22 &  RN91 \\
        & 4800 & -0.05 &  &  \nodata  & -0.03 &  Reid01 \\
HIP 74235  & 4900 & -1.51 &  & 5040 & -1.54 &  TL99 \\
        & 4900 & -1.51 &  &  \nodata  & -1.31 &  TI99 \\
        & 4900 & -1.51 &  & 4957 & -1.57 &  CLLA94 \\
        & 4900 & -1.51 &  & 4974 & -1.52 &  AAM96 \\
        & 4900 & -1.51 &  & 4850 & -1.40 &  Fulbright00 \\
BD +2 2944  & 6000 & -0.02 &  & 6036 & -0.13 &  AAM99 \\
G 17-25  & 4950 & -1.46 &  & 5175 & -1.10 &  Fulbright00 \\
        & 4950 & -1.46 &  &  \nodata  & -1.08 &  TI99 \\
        & 4950 & -1.46 &  & 5180 & -1.35 &  TL99 \\
        & 4950 & -1.46 &  & 4966 & -1.34 &  AAM96 \\
        & 4950 & -1.46 &  & 4959 & -1.54 &  CLLA94 \\
LP 445-55   & 5200 & -1.80 &  & \nodata & -1.55 &  RN91 \\
LP 447-2    & 5800 & -1.79 &  & \nodata & -1.74 &  RN91 \\
LP 747-18   & 5800 & -0.95 &  & \nodata & -1.22 &  RN91 \\
G 19-25  & 4900 & -2.01 &  &  \nodata  & -2.03 &  RN91 \\
        & 4900 & -2.01 &  & 4938 & -1.88 &  CLLA94 \\
        & 4900 & -2.01 &  & 4977 & -1.52 &  AAM96 \\
BD -8 4501  & 5800 & -1.80 &  & 5657 & -1.99 &  CLLA94 \\
        & 5800 & -1.80 &  & 5682 & -2.79 &  AAM96 \\
G 204-47    & 5200 & -0.88 &  & 5249 & -0.86 &  CLLA94 \\
BD +5 3640  & 4950 & -1.34 &  & 5080 & -1.27 &  TL99 \\
        & 4950 & -1.34 &  & 5012 & -1.36 &  CLLA94 \\
        & 4950 & -1.34 &  & 4980 & -1.44 &  AAM96 \\
        & 4950 & -1.34 &  &  \nodata  & -0.37 &  RN91 \\
G 227-44    & 5400 & -1.45 &  & 5501 & -1.41 &  CLLA94 \\
G 207-23    & 5000 & -1.90 &  & \nodata & -1.00 &  RWMRKM01  \\
HD 181007   & 4700 & -1.93 &  & 4582 & -2.27 &  AAM96\tablenotemark{a}  \\
LP 752-18   & 5800 & -1.00 &  & \nodata & -1.73 &  RN91  \\
LP 753-29   & 5400 & -1.88 &  & \nodata & -2.10 &  RN91 \\
G 23-14     & 5000 & -1.51 &  & 5100 & -1.40 &  Carney97 \\
        & 5000 & -1.51 &  & 4893 & -2.20 &  CLLA94 \\
LP 815-43   & 6400 & -2.70 &  & 6379 & -2.64 &  TI99 \\
        & 6400 & -2.70 &  & 6500 & -2.95 &  PMBH00 \\
BD +22 4567  & 4800 & -0.33 &  &  \nodata  & -0.22 &  Eggen98 \\
HIP 109801  & 4600 & -1.70 &  & 4661 & -1.43 &  CLLA94 \\
        & 4600 & -1.70 &  &  \nodata  & -0.92 &  RN91 \\
ROSS 237  & 4900 & -1.49 &  & 4947 & -1.38 &  CLLA94 \\
BD +28 4634  & 5000 & -0.30 &  &  \nodata  & -0.15 &  SPC93 \\
\enddata 

\tablenotetext{a}{Giant star assumed to be a dwarf}

\tablecomments{
The following references were based on spectroscopic
data: Beers99, CLLA94, Carney97, Clementini99, Fulbright00,
Gizis97, Primas00, RN91, TI99, TL99.
}

\tablerefs{
(AAM96) = \citet{alonso96};
(AAM99) = \citet{alonso99};
(Beers99) = \citet{beers99};
(CLLA94) = \citet{carney94};
(Carney97) = \citet{carney97};
(Clementini99) = \citet{clementini99};
(Eggen87) = \citet{eggen87};
(Eggen98) = \citet{eggen98};
(Fulbright00) = \citet{fulbright00};
(GL00) = \citet{gl2000};
(Gizis97) = \citet{gizis97};
(Morrison01) = \citet{morrison01};
(Norris86) = \citet{norris86};
(Primas00) = \citet{primas00};
(Reid01) = \citet{reid01};
(RN91) = \citet{ryan91};
(SPC93) = \citet{schuster93};
(TI99) = \citet{thevenin99};
(TL99) = \citet{tomkin99}
}

\end{deluxetable} 

%% file: tab3.tex
\begin{deluxetable}{lccrc} 
\tabletypesize{\tiny}
\tablecolumns{7} 
\tablewidth{0pc} 
\tablecaption{Comparison of \teff's derived in this study
with \teff's calculated from the \citet{alonso96b,alonso99b}
relations for {\em \teff:[Fe/H]:color} using photometry
found in the literature.  
\label{irfm}}
\tablehead{ 
\colhead{Star}       & \colhead {\teff~(K)}  &
\colhead{[Fe/H]}     & \colhead {Color}&
\colhead{\teff~(K) from}      \\
\colhead{}           & \multicolumn{2}{c}{This study} &
                       \colhead{Index} &
\colhead{color relation}
}
\startdata 
PLX 5805 & 4600 & -1.72 & B-V & 4764 \\
    & 4600 & -1.72 & J-K & 4464 \\
BD -14 142  & 6200 & -1.21 & B-V & 5113 \\
    & 6200 & -1.21 & V-K & 6327 \\
    & 6200 & -1.21 & J-K & 5860 \\
LP 646-24  & 4800 & -1.01 & V-K & 4666 \\
    & 4800 & -1.01 & J-K & 4665 \\
G 72-34   & 4800 & -2.23 & J-K & 4654 \\
BD -4 290  & 4900 & -0.01 & B-V & 4544 \\
BD +8 335  & 5100 & -0.98 & B-V & 4485 \\
HIP 10412  & 4600 & -0.40 & B-V & 4491 \\
Cool 340  & 4800 & -0.40 & B-V & 4581 \\
    & 4800 & -0.40 & V-K & 4158 \\
    & 4800 & -0.40 & J-K & 4544 \\
BD +4 415  & 4750 & -0.63 & AAM96\tablenotemark{a} & 4756 \\
GJ 1064A  & 5000 & -1.15 & B-V & 4717 \\
    & 5000 & -1.15 & V-K & 4963 \\
    & 5000 & -1.15 & J-K & 4978 \\
GJ 1064B  & 4800 & -1.15 & B-V & 4692 \\
    & 4800 & -1.15 & V-K & 4591 \\
    & 4800 & -1.15 & J-K & 5791 \\
BD -4 680  & 5800 & -1.89 & AAM96\tablenotemark{a} & 5866 \\
G 39-36  & 4200 & -2.50 & J-K & 4400 \\
G 81-41  & 4550 & -0.96 & B-V & 4357 \\
BD -10 1085  & 5100 & -0.50 & B-V & 4676 \\
    & 5100 & -0.50 & V-K & 4982 \\
    & 5100 & -0.50 & J-K & 4978 \\
BD +19 869  & 4600 & -1.01 & V-K & 4245 \\
    & 4600 & -1.01 & J-K & 4287 \\
PLX 1219  & 4600 & -1.79 & AAM96\tablenotemark{a} & 4739 \\
HIP 27928  & 4200 & -0.93 & J-K & 4082 \\
G 103-27  & 5600 & -0.95 & B-V & 5580 \\
    & 5600 & -0.95 & V-K & 5118 \\
    & 5600 & -0.95 & J-K & 5153 \\
G 101-34  & 5000 & -1.50 & B-V & 4724\tablenotemark{b} \\
G 103-50 & 4700 & -2.20 & AAM96\tablenotemark{a} & 4713 \\
HIP 30567  & 4700 & -0.60 & B-V & 4701 \\
AC +78 2199  & 5000 & -0.03 & B-V & 4408 \\
    & 5000 & -0.03 & V-K & 5036 \\
    & 5000 & -0.03 & J-K & 5026 \\
G 88-1  & 4600 & -0.84 & B-V & 4700 \\
    & 4600 & -0.84 & V-K & 4394 \\
    & 4600 & -0.84 & J-K & 4482 \\
BD -17 1716  & 4600 & -0.08 & B-V & 4438 \\
    & 4600 & -0.08 & V-K & 4397 \\
    & 4600 & -0.08 & J-K & 4301 \\
G 87-27  & 5100 & -0.61 & B-V & 4697\tablenotemark{b} \\
G 87-35  & 5300 & -0.53 & B-V & 6191 \\
    & 5300 & -0.53 & V-K & 5055 \\
    & 5300 & -0.53 & J-K & 5242 \\
G 193-49  & 5000 & -0.39 & B-V & 4538 \\
G 90-4  & 5600 & -0.81 & b-y,c1 & 5473 \\
LP 208-60  & 5800 & -0.22 & J-K & 5121 \\
G 115-1  & 4800 & -1.33 & B-V & 4407 \\
    & 4800 & -1.33 & V-K & 4884 \\
    & 4800 & -1.33 & J-K & 4784 \\
PLX 2019  & 4600 & -1.72 & AAM96\tablenotemark{a} & 4740 \\
LP 209-14  & 5400 & -1.36 & J-K & 5701 \\
G 9-31  & 5600 & -1.00 & AAM96\tablenotemark{a} & 5569 \\
LP 36-78  & 5150 & 0.25 & B-V & 4786 \\
LP 786-61  & 5800 & -0.76 & B-V & 5557 \\
ROSS 94  & 4600 & 0.00 & B-V & 3652\tablenotemark{b} \\
    & 4600 & 0.00 & V-K & 3744\tablenotemark{b} \\
    & 4600 & 0.00 & J-K & 4207\tablenotemark{b} \\
G 116-55  & 5600 & -2.10 & B-V & 5513 \\
    & 5600 & -2.10 & V-K & 5740 \\
    & 5600 & -2.10 & J-K & 5486 \\
G 43-7  & 4950 & -1.02 & B-V & 6627 \\
G 161-84  & 4700 & -1.31 & J-K & 4471 \\
LP 788-55  & 4700 & -0.53 & J-K & 4612 \\
WOLF 334  & 5200 & -1.71 & J-K & 5455 \\
LP 315-12  & 5100 & -0.98 & V-K & 4582 \\
    & 5100 & -0.98 & J-K & 4756 \\
LP 429-17  & 4750 & -0.10 & B-V & 4658 \\
    & 4750 & -0.10 & V-K & 4563 \\
    & 4750 & -0.10 & J-K & 4604 \\
BD +12 2201  & 4500 & -0.23 & V-K & 4245 \\
    & 4500 & -0.23 & J-K & 4095 \\
LP 790-19  & 4300 & -0.51 & J-K & 4119 \\
BD -9 3104  & 4900 & -0.60 & B-V & 4853 \\
    & 4900 & -0.60 & V-K & 4785 \\
    & 4900 & -0.60 & J-K & 4884 \\
G 119-21  & 4600 & -0.48 & B-V & 4126 \\
    & 4600 & -0.48 & V-K & 4304 \\
    & 4600 & -0.48 & J-K & 4604 \\
BD +31 2175  & 4800 & -0.17 & B-V & 4502 \\
    & 4800 & -0.17 & V-K & 4826 \\
    & 4800 & -0.17 & J-K & 4673 \\
PLX 2529.1  & 4800 & -0.58 & B-V & 4610 \\
    & 4800 & -0.58 & V-K & 4653 \\
    & 4800 & -0.58 & J-K & 4650 \\
G 119-45  & 4800 & 0.14 & J-K & 4537 \\
G 58-36  & 4600 & -0.05 & J-K & 4540 \\
G 253-46  & 5200 & -0.57 & V-K & 5300 \\
    & 5200 & -0.57 & J-K & 5394 \\
HIP 55128  & 4900 & -0.38 & B-V & 4609 \\
    & 4900 & -0.38 & V-K & 4685 \\
    & 4900 & -0.38 & J-K & 4728 \\
LP 792-12  & 4800 & -0.39 & B-V & 4261 \\
G 122-22  & 4400 & -1.36 & V-K & 4465 \\
G 163-80  & 4700 & 0.05 & J-K & 4658 \\
ROSS 109  & 5300 & -1.53 & B-V & 5610 \\
LP 733-14  & 4700 & -0.57 & V-K & 4534 \\
    & 4700 & -0.57 & J-K & 4407 \\
WOLF 1424  & 4600 & -1.19 & V-K & 4491 \\
    & 4600 & -1.19 & J-K & 4358 \\
LP 734-101  & 5500 & 0.06 & B-V & 5548 \\
G 13-29  & 4800 & -0.66 & V-K & 4684 \\
    & 4800 & -0.66 & J-K & 4338 \\
G 198-61  & 5300 & -0.12 & B-V & 5147 \\
    & 5300 & -0.12 & V-K & 5139 \\
    & 5300 & -0.12 & J-K & 5460 \\
G 149-9  & 4900 & -0.62 & V-K & 4731 \\
    & 4900 & -0.62 & J-K & 4631 \\
HIP 62627  & 4800 & -0.24 & B-V & 4295 \\
G 149-18  & 5000 & -0.03 & B-V & 4206 \\
    & 5000 & -0.03 & V-K & 4643 \\
    & 5000 & -0.03 & J-K & 4650 \\
HD 114095  & 4730 & -0.71 & B-V & 4422\tablenotemark{b} \\
    & 4730 & -0.71 & V-K & 4510\tablenotemark{b} \\
    & 4730 & -0.71 & J-K & 4565\tablenotemark{b} \\
    & 4730 & -0.71 & AAM96\tablenotemark{a} & 4541\tablenotemark{c} \\
BD +68 714  & 4900 & -0.05 & B-V & 4424 \\
    & 4900 & -0.05 & V-K & 4797 \\
    & 4900 & -0.05 & J-K & 4952 \\
ROSS 466  & 5000 & -0.86 & J-K & 4784 \\
LHS 2715  & 4400 & -1.56 & AAM96\tablenotemark{a} & 4480 \\
G 65-22 & 5000 & -1.66 & AAM96\tablenotemark{a} & 4971 \\
LP 172-89  & 4800 & -0.49 & B-V & 4384 \\
LP 738-45  & 4400 & -0.12 & J-K & 4264 \\
G 255-44  & 4900 & -1.30 & B-V & 4901 \\
    & 4900 & -1.30 & V-K & 4686 \\
    & 4900 & -1.30 & J-K & 4813 \\
BD +30 2490  & 4600 & -0.18 & B-V & 4005 \\
    & 4600 & -0.18 & V-K & 4185 \\
    & 4600 & -0.18 & J-K & 4442 \\
G 239-12  & 5800 & -2.62 & B-V & 5460 \\
HIP 70152  & 4800 & -0.05 & B-V & 4446 \\
LP 500-92  & 4500 & -0.38 & B-V & 3765 \\
    & 4500 & -0.38 & V-K & 4184 \\
    & 4500 & -0.38 & J-K & 4294 \\
LP 175-8  & 4900 & -0.31 & B-V & 4490 \\
    & 4900 & -0.31 & V-K & 4678 \\
    & 4900 & -0.31 & J-K & 4842 \\
HD 131287  & 4800 & -0.04 & B-V & 5715 \\
LP 502-8  & 5000 & -0.95 & B-V & 4559 \\
HIP 74235  & 4900 & -1.51 & AAM96\tablenotemark{a} & 4971 \\
BD +2 2944  & 6000 & -0.02 & AAM99\tablenotemark{a} & 5856\tablenotemark{b} \\
G 225-50  & 4600 & -0.13 & B-V & 3964 \\
    & 4600 & -0.13 & V-K & 4637 \\
    & 4600 & -0.13 & J-K & 4341 \\
LP 330-7  & 4800 & -0.22 & B-V & 4087 \\
    & 4800 & -0.22 & V-K & 4743 \\
    & 4800 & -0.22 & J-K & 4708 \\
G 202-57  & 4900 & -1.05 & B-V & 4917 \\
    & 4900 & -1.05 & V-K & 4617 \\
    & 4900 & -1.05 & J-K & 4612 \\
G 17-25  & 4950 & -1.46 & AAM96\tablenotemark{a} & 4966 \\
BD -14 4454  & 4600 & -0.95 & V-K & 4474 \\
    & 4600 & -0.95 & J-K & 4334 \\
LP 445-55  & 5200 & -1.80 & V-K & 5807 \\
    & 5200 & -1.80 & J-K & 5420 \\
LP 447-2  & 5800 & -1.79 & B-V & 5404 \\
    & 5800 & -1.79 & V-K & 5611 \\
    & 5800 & -1.79 & J-K & 5491 \\
LP 747-18  & 5800 & -0.95 & B-V & 5380 \\
    & 5800 & -0.95 & V-K & 4827 \\
    & 5800 & -0.95 & J-K & 5103 \\
LP 447-34  & 4800 & -0.35 & B-V & 5136 \\
    & 4800 & -0.35 & V-K & 4699 \\
    & 4800 & -0.35 & J-K & 4760 \\
G 19-25 & 4900 & -2.01 & AAM96\tablenotemark{a} & 4977 \\
BD -8 4501  & 5800 & -1.80 & AAM96\tablenotemark{a} & 5682 \\
G 204-47  & 5200 & -0.88 & B-V & 4923 \\
    & 5200 & -0.88 & V-K & 5044 \\
    & 5200 & -0.88 & J-K & 5394 \\
BD +5 3640 & 4950 & -1.34 & AAM96\tablenotemark{a} & 4980 \\
BD -17 5287  & 5800 & -0.70 & B-V & 4584 \\
G 227-44  & 5400 & -1.45 & B-V & 4932 \\
G 207-23  & 5000 & -1.90 & B-V & 4798 \\
    & 5000 & -1.90 & V-K & 5279 \\
    & 5000 & -1.90 & J-K & 5176 \\
HD 181007  & 4700 & -1.93 & B-V & 4659\tablenotemark{b} \\
    & 4700 & -1.93 & V-K & 4676\tablenotemark{b} \\
    & 4700 & -1.93 & J-K & 4887\tablenotemark{b} \\
LP 752-18  & 5800 & -1.00 & B-V & 6055 \\
G 23-14  & 5000 & -1.51 & b-y  & 4958\tablenotemark{b} \\
LP 575-39  & 4700 & -0.42 & B-V & 4100 \\
LP 575-40  & 4800 & -0.10 & B-V & 4349 \\
LP 815-43  & 6400 & -2.70 & B-V & 5193 \\
    & 6400 & -2.70 & V-K & 6509 \\
    & 6400 & -2.70 & J-K & 5973 \\
HD 200968  & 5000 & -0.16 & B-V & 4550 \\
BD +22 4567  & 4800 & -0.33 & B-V & 4315 \\
BD +30 4633  & 4600 & -0.01 & B-V & 3889 \\
HIP 109801  & 4600 & -1.70 & J-K & 4593 \\
ROSS 237  & 4900 & -1.49 & B-V & 5410 \\
    & 4900 & -1.49 & V-K & 4590 \\
    & 4900 & -1.49 & J-K & 4879 \\
ROSS 242  & 4700 & -0.94 & B-V & 4680 \\
    & 4700 & -0.94 & V-K & 4682 \\
    & 4700 & -0.94 & J-K & 4646 \\
LP 702-79  & 4800 & 0.05 & V-K & 4605 \\
    & 4800 & 0.05 & J-K & 4331 \\
HIP 115664  & 4600 & -0.90 & V-K & 4370 \\
    & 4600 & -0.90 & J-K & 4411 \\
BD +28 4634  & 5000 & -0.30 & B-V & 4515 \\
    & 5000 & -0.30 & V-K & 4825 \\
    & 5000 & -0.30 & J-K & 4712 \\
\enddata 

\tablenotetext{a}{\teff's determined from direct application of the IRFM}
\tablenotetext{b}{Giant star; \teff~derived from \citet{alonso99b} relation}
\tablenotetext{c}{Giant star assumed to be a dwarf}

\tablecomments{
Photometry from the following sources.
B and V taken from the Guide Star Catalog II.
J and K taken from the Two Micron All Sky Survey (2MASS) \citet{2mass}.
b-y and c1 taken from \citet{schuster93}.
}

\tablerefs{
(AAM96) = \citet{alonso96};
(AAM99) = \citet{alonso99}
}

\end{deluxetable}